\documentclass[a4paper,10pt,twocolumn]{elsarticle}

\usepackage{float}
\usepackage{amsmath}
\usepackage{amssymb}
\usepackage{graphicx}
\usepackage{threeparttable}
\usepackage{eurosym}
\usepackage{array}
\usepackage{url}
\usepackage[top=3em, bottom=5em, left=3em, right=3em]{geometry}

\usepackage{comment}
\usepackage[table,xcdraw]{xcolor}
\usepackage{ulem}

\usepackage{fancyhdr}
\def \kIndex {k}
\def \pIndex {p}

\def \iIndex {i}

\def \dkNonPUN {d^\zeta_{t\kIndex}}
\def \dkPUNw {d^w_{t\kIndex}}
\def \dkPUNd {d^d_{t\kIndex}}

\def \pk {P^d_{t\kIndex}}
\def \ph {P^d_{th}}
\def \sp {s_{t\pIndex}}
\def \pp {P^s_{t\pIndex}}

\def \vphikNonPUN {\varphi^\zeta_{t\kIndex}}
\def \vphikPUNw {\varphi^w_{t\kIndex}}
\def \vphikPUNwlo {\varphi^{w,lo}_{t\kIndex}}
\def \vphip {\varphi^s_{t\pIndex}}

\def \deltaMax {\delta_{tij}^{max}}
\def \etaij {\eta_{tij}}
\def \etaji {\eta_{tji}}

\def \Fij {f_{tij}}
\def \Fji {f_{tji}}
\def \Fijmax {F_{tij}^{max}}

\def \zAll    {\mathcal{Z}}
\def \zNonPUN {\mathcal{Z}^\zeta}
\def \zPUN    {\mathcal{Z}^\pi}

\def \zI    {\iIndex}
\def \zJ    {j}

\def \kSetPUN {\mathcal{K}^\pi_t}
\def \kSet {\mathcal{K}_t}
\def \kSetNonPUN {\mathcal{K}^\zeta_t}

\def \kSetPUNi {\mathcal{K}^\pi_{t\iIndex}}
\def \kSeti {\mathcal{K}_{t\iIndex}}
\def \kSetNonPUNi {\mathcal{K}^\zeta_{t\iIndex}}

\def \pSet {\mathcal{P}_t}
\def \pSeti {\mathcal{P}_{t\iIndex}}

\def \pZi {\zeta_{t\iIndex}}

\def \dkmax {D_{t\kIndex}^{max}}
\def \spmax {S_{tp}^{max}}

\def \ugk {u^g_{t\kIndex}}
\def \ugh {u^g_{th}}
\def \uek {u^e_{t\kIndex}}

\def \uwk {u^w_{t\kIndex}}
\def \udk {u^d_{t\kIndex}}

\def \uf {u^{f}_{tij}}

\def \bji {b_{tji}}

\def \meritk {O^m_{tk}}
\def \merith {O^m_{th}}

\def \yugPUNk {y^{g \pi}_{tk}}
\def \yugPzk  {y^{g \zeta}_{tki}}

\def \yuePUNk {y^{e \pi}_{tk}}
\def \yuwvphik {y^{w \varphi}_{tk}}

\def \yubVphibMax {y^{B \varphi, max}_{p}}
\def \yubVphibMin {y^{B \varphi, min}_{p}}

\def \ybPzi {y^{b\zeta}_{tij}}

\def \dkpi {d^\pi_{tk}}

\def \tSet {\mathcal{T}}
\def \tSetBlockp {\mathcal{T}_p}

\def \pSetBlocki {\mathcal{P}^B_i}
\def \pSetBlock {\mathcal{P}^B}

\def \ppb {P^B_{\pIndex}}

\def \spbmax {S_{tp}^{B, max}}

\def \ubp {u^B_p}
\def \mar {R^{min}_p}
\def \rblockAccepted {r_p}

\def \vphibMax {\varphi^{B, max}_{\pIndex}}
\def \vphibMin {\varphi^{B, min}_{\pIndex}}

\def \imbalance {\kappa_t}

\def \bigMpi {M^\pi}

\def \bigMFlow {M^F_{tij}}
\def \bigMDmax {M^D_{th}}

\def \bigMBlock {M^B_p}

\def \mycolor {black}

\begin{document}

\onecolumn

\begin{frontmatter}

\title{A New Approach to Electricity Market Clearing With Uniform Purchase Price and Curtailable Block Orders}

\author[siena]{Iacopo~Savelli\corref{cor1}}
\ead{savelli@diism.unisi.it}

\author[montefiore]{Bertrand~Corn\'elusse}
\ead{bertrand.cornelusse@uliege.be}

\author[siena]{Antonio~Giannitrapani}
\ead{giannitrapani@diism.unisi.it}

\author[siena]{Simone~Paoletti}
\ead{paoletti@diism.unisi.it}

\author[siena]{Antonio~Vicino}
\ead{vicino@diism.unisi.it}

\cortext[cor1]{Corresponding author}

\address[siena]{Dipartimento di Ingegneria dell'Informazione e Scienze Matematiche,
Universit\`a di Siena, Siena 53100, Italy}

\address[montefiore]{Department of Electrical Engineering and Computer Science,
University of Li\`ege, Belgium}

\begin{abstract}
The European market clearing problem is characterized by a set of heterogeneous orders and rules that force the implementation of heuristic and iterative solving methods. In particular, curtailable block orders and the uniform purchase price pose serious difficulties. A block order spans over multiple hours, and can be either fully accepted or fully rejected. The uniform purchase price prescribes that all consumers pay a common price in all the zones, while producers receive zonal prices, which can differ from one zone to another.

The market clearing problem in the presence of both the uniform purchase price and block orders is a major open issue in the European context. The uniform purchase price scheme leads to a non-linear optimization problem involving both primal and dual variables, whereas block orders introduce multi-temporal constraints and binary variables into the problem. As a consequence, the market clearing problem in the presence of both block orders and the uniform purchase price can be regarded as a non-linear integer programming problem involving both primal and dual variables with complementary and multi-temporal constraints.

The aim of this paper is to present a non-iterative and heuristic-free approach for solving the market clearing problem in the presence of both curtailable block orders and the uniform purchase price scheme. The solution is exact, with no approximation up to the level of resolution of current market data. By resorting to an equivalent uniform purchase price formulation, the proposed approach results in a mixed-integer linear program, which is built starting from a non-linear integer bilevel programming problem. Numerical results using real market data are reported to show the effectiveness of the proposed approach. The model has been implemented in Python, and the code is freely available on a public repository.
\end{abstract}


\end{frontmatter}
\twocolumn

\section*{Nomenclature}
\begin{table}[!h]
\normalsize
\setlength{\extrarowheight}{1pt}
\setlength{\tabcolsep}{6pt}
\begin{tabular}{lp{0.85\columnwidth}}
\multicolumn{2}{l}{ \textit{A. Sets and Indices}}\\
$\zI$      & Index of market zones, $\zI \in \zAll$.\\
$\kSetPUNi$    & Set of consumers paying the UPP $\pi_t$ in zone $\zI \in \zAll$, with $t \in \tSet$.\\
$\kSetPUN$     & Set of all consumers paying the UPP, i.e., $\kSetPUN = \cup_i \kSetPUNi$, with $t \in \tSet$. \\
$\kSetNonPUNi$ & Set of consumers paying the zonal price $\pZi$ in zone $\zI \in \zAll$, with $t \in \tSet$.\\ $\kSetNonPUN$  & Set of all consumers paying zonal prices, i.e., $\kSetNonPUN = \cup_i \kSetNonPUNi$, with $t \in \tSet$.\\
$\kSeti$       & Set of all consumers in zone $\zI \in \zAll$, i.e., ${\kSeti=\kSetPUNi \cup \kSetNonPUNi}$, with $t \in \tSet$.\\
$\kSet$        & Set of all consumers, i.e., $\kSet = \cup_i \kSeti$, with $t \in \tSet$.\\
$\pSeti$       & Set of producers submitting simple stepwise order in zone $\zI\,{\in \zAll}$, with $t\,{\in \tSet}$.\\
$\pSet$        & Set of all producers submitting simple stepwise order, i.e., $\pSet = \cup_i \pSeti$, $t \in \tSet$.\\
$\pSetBlocki$  & Set of producers submitting curtailable profile block orders in zone $\zI \in \zAll$.\\
$\pSetBlock$   & Set of all producers submitting curtailable profile block orders, i.e., $\pSetBlock = \cup_i \pSetBlocki$.\\
$\tSet$        & Set of the 24 daily hours.\\
$\tSetBlockp$  & Set of block order $p$ timespan, with $\pIndex \in \pSetBlock$, and $\tSetBlockp \subseteq \tSet$.\\
$\zPUN$    & Set of zones enforcing the UPP $\pi_t$.\\
$\zNonPUN$ & Set of zones without the UPP, all consumers pay zonal prices $\pZi$.\\
$\zAll$    & Set of all zones, $\zAll = \zPUN \cup \zNonPUN$.\\
\end{tabular}
\end{table}

\begin{table}[!h]
\normalsize
\setlength{\extrarowheight}{1pt}
\setlength{\tabcolsep}{1pt}
\begin{tabular}{lp{0.85\columnwidth}}
\multicolumn{2}{l}{ \textit{B. Constants}}\\
$\dkmax$  & Maximum hourly quantity demanded by consumer $\kIndex \in \kSet$ at time $t \in \tSet$.\\
$\Fijmax$ & Maximum flow capacity from zone $\zI$ to zone $\zJ$ with $t \in \tSet$.\\
$M^{(.)}$  &   \textcolor{\mycolor}{$\bigMpi$, $\bigMBlock$, $\bigMFlow$, and $\bigMDmax$ are big-M parameters.}\\
$\meritk$& Merit order for consumer $\kIndex\,{\in \kSetPUN}$, lower values mean higher priority.\\
$\pk$     & Hourly order price submitted by consumer $\kIndex~{\in \kSet}$ with $t \in \tSet$.\\
$\pp$     & Hourly order price submitted by producer ${\pIndex \in \pSet}$ with $t \in \tSet$.\\
$\ppb $   & Block order price submitted by producer ${\pIndex \in \pSetBlock}$.\\
$\mar$ & Minimum acceptance ratio for curtailable block order with $\pIndex~{\in\pSetBlock}$.\\
$\spmax$  & Maximum hourly quantity offered by producer $\pIndex \in \pSet$ at time $t \in \tSet$.\\
$\spbmax$ & Profile block order maximum hourly quantity offered by producer $\pIndex \in \pSetBlock$ with $t \in \tSetBlockp$.\\
\end{tabular}
\end{table}

\begin{table}[!h]
\normalsize
\setlength{\extrarowheight}{1pt}
\setlength{\tabcolsep}{1pt}
\begin{tabular}{lp{0.85\columnwidth}}
\multicolumn{2}{l}{ \textit{C. Variables}}\\
$\bji$ & Binary variable used in the binary expansion to convert a positive integer number in binary form, where $i \in \zPUN$, $t \in \tSet$, and $j \in \{0, \ldots\}$.\\
$\dkNonPUN$ & Executed demand quantity for consumer $\kIndex \in \kSetNonPUN$, with $t \in \tSet$.\\
$\dkPUNw$ & Executed demand quantity for consumer $\kIndex \in \kSetPUN$ if $\uwk=1$, with $t \in \tSet$.\\
$\dkPUNd$ & Executed demand quantity for consumer $\kIndex \in \kSetPUN$ if $\udk=1$, with $t \in \tSet$.\\
$\dkpi$ &  Executed demand quantity for consumer $\kIndex \in \kSetPUN$, where $\dkpi = \ugk \dkmax + \dkPUNw + \dkPUNd$, with $\kIndex \in \kSetPUN$ and $t \in \tSet$.\\
$\Fij$ & Flow from zone $\zI$ to zone $\zJ$  with $t \in \tSet$.\\
$\rblockAccepted$ & Block order acceptance ratio with $\pIndex \in \pSetBlock$.\\
$\sp$ & Executed supply quantity for producer $\pIndex \in \pSet$, with $t \in \tSet$.\\
$\uf$ & Binary variable with $i, j \in \zPUN$ and $t \in \tSet$, where $\uf=1$ if and only if the transmission line from $i$ to $j$ is congested, i.e., $\Fij = \Fijmax$.\\
$\ubp$ & Binary variable representing the block order acceptance status with $\pIndex \in \pSetBlock$, where $\ubp=1$ means accepted, and $\ubp=0$  rejected.\\
$\ugk$ & Binary variable with $\kIndex \in \kSetPUN$ and $t \in \tSet$, where $\ugk~{=1}~{\iff\pk}~{> \pi}$, and zero otherwise.\\
$\uek$ & Binary variable with $\kIndex \in \kSetPUN$ and $t \in \tSet$, where if $\uek=1$ then $\pk = \pi$.\\
$\uwk$ & Binary variable with $\kIndex \in \kSetPUN$ and $t \in \tSet$, where if $\uwk=1$ then the demand order is at-the-money and it is partially cleared according to a social welfare approach.\\
$\udk$ & Binary variable with $\kIndex \in \kSetPUN$ and $t \in \tSet$, where if $\udk=1$ then the demand order is at-the-money and it is partially cleared according to an economic dispatch approach.\\
$\deltaMax$ & Dual variable of constraint $\Fij \leq \Fijmax$.\\
$\pZi$ & Zonal price in zone $\zI \in \zAll$  with $t \in \tSet$.\\
$\etaij$ & Dual variable of constraint $\Fij + \Fji = 0$.\\
$\imbalance$ & Error tolerance in the uniform purchase price definition, currently $\imbalance \in [-1;5]$.\\
$\vphikNonPUN$ & Dual variable of constraint $\dkNonPUN \leq \dkmax$.\\
$\vphikPUNw$ & Dual variable of constraint $\dkPUNw \leq \dkmax\uwk$.\\
$\vphikPUNwlo$ & Dual variable of constraint $\dkPUNw \geq 0$.\\
$\vphip$ & Dual variable of constraint $\sp \leq \spmax$.\\
$\vphibMax $ & Dual variable of constraint $\rblockAccepted \leq \ubp$.\\
$\vphibMin $ & Dual variable of constraint $\rblockAccepted \geq \ubp\mar$.\\
$\pi_t$ & Uniform purchase price  at time $t \in \tSet$.\\
\end{tabular}
\end{table}

\begin{table}[!h]
\normalsize
\setlength{\extrarowheight}{1pt}
\setlength{\tabcolsep}{1pt}
	\begin{tabular}{lp{0.85\columnwidth}}
		\multicolumn{2}{l}{{\normalsize \textit{D. Auxiliary Variables}}}\\
		$\yugPUNk$       &  Auxiliary variable, it replaces the product $\ugk \pi$. \\
		$\yugPzk $       &  Auxiliary variable, it replaces the product $\ugk \pZi$. \\
		$\yuePUNk$       &  Auxiliary variable, it replaces the product $\uek \pi_t$. \\
		$\yuwvphik$      &  Auxiliary variable, it replaces the product $\uwk \vphikPUNw$. \\
		$\yubVphibMax$ 	 &  Auxiliary variable, it replaces the product $\ubp \vphibMax$. \\
		$\yubVphibMin$	 &  Auxiliary variable, it replaces the product $\ubp \vphibMin$. \\
		$\ybPzi$         &  Auxiliary variable, it replaces the product $\bji \pZi$. \\		
	\end{tabular}
\end{table}

\section{Introduction}

\textcolor{\mycolor}{Electricity markets are experiencing significant changes due to different factors, as the modification of generation mix \cite{ZOU201756}, the increasing presence of demand response \cite{MARKLEHU20181290} and energy storage systems \cite{KHAN201839}, the growth of renewable energy \cite{IYCHETTIRA2017228}, the request for both flexibility \cite{OLIVELLAROSELL2018881, AYON20171} and security of supply \cite{DUENAS2018443}, and the associated adjustment in power networks \cite{Forte2017}. These changes affected also the European markets.}
In particular, the current day-ahead European electricity market is the result of a merging process that took place during the last three decades and involved all the main European countries \cite{glachant2016mapping}, and it should lead to significant social welfare improvements \cite{strbac016benefitsofintegrating}. However, the complete integration involves several difficulties both in terms of design \cite{interactionBetweenMarkets2017}, and interaction between different markets \cite{balancingMarketNorth}. In particular, the lack of an original common design, leads to a European day-ahead electricity market that is characterized by heterogeneous orders (e.g., stepwise orders, piecewise linear orders, simple and linked block orders \cite{EUPHEMIAmanual}), and rules (e.g., minimum income condition \cite{OMIEsite}, uniform purchase price \cite{sitoGME}), which cannot be easily harmonized. As a consequence, the European market clearing algorithm \cite{EUPHEMIAmanual} deals with a wide variety of issues, due to, for example, the complexity of both the clearing rules and the orders involved, their heterogeneous nature, and the increasing number of orders currently submitted to the market, which forced the implementation of heuristics and iterative solving methods. One of the most challenging problem is the simultaneous presence of block orders and the uniform purchase pricing scheme. Block orders are present in the central and northern European countries \cite{EPEXsite,NordPoosite}, whereas the uniform purchase price (UPP) is implemented into the Italian market \cite{sitoGME} with the name of \textit{Prezzo Unico Nazionale} (PUN).

\subsection{The UPP scheme}\label{UPP_section_Intro}

The UPP scheme requires that all consumers pay a unique price, termed the UPP, in all the zones, while producers receive zonal prices, which can differ from one zone to another. The UPP $\pi_t$ at time $t$ is defined as the average of the zonal prices $\pZi$, weighted by the consumers' cleared quantities $\dkpi$. Formally:
\begin{equation}\label{PUNdef}
\pi_t = \dfrac{\sum_{i \in \zPUN} \sum_{\kIndex \in \kSetPUNi}  \pZi \dkpi}{\sum_{\kIndex \in \kSetPUN} \dkpi}\,,
\end{equation}
where $\zPUN$ is the set of zones enforcing the UPP, $\kSetPUNi$ is the set of consumers paying the UPP in the zone $\iIndex$ at time $t \in \tSet$, and $\kSetPUN=\cup_i \kSetPUNi$.

Given the UPP definition \eqref{PUNdef}, it is possible to specify the following \textit{UPP clearing rule}:
\begin{itemize}
	\item demand orders with a submitted price $\pk$ strictly greater than $\pi_t$, that is, in-the-money (ITM) demand orders, must be fully executed, i.e., ${\dkpi=\dkmax}$;
	\item demand orders with a submitted price $\pk$ exactly equal to $\pi_t$, that is, at-the-money (ATM) demand orders, may be partially cleared, i.e., ${0\leq\dkpi\leq\dkmax}$;
	\item demand orders with a submitted price $\pk$ strictly lower than $\pi_t$, that is, out-of-the-money (OTM) demand orders, must be fully rejected, i.e., ${\dkpi=0}$.
\end{itemize}

In addition, demand orders subject to the UPP scheme are ranked by a parameter termed \textit{merit order}, that determines a strict total ordering between the UPP orders. The merit order is assigned by the market operator before the day-ahead auction. Lower merit order $\meritk$ implies a higher priority in execution. In particular, this ranking coincides with the price ranking for the orders with different submitted prices, i.e., if ${P^d_{th} > \pk}$ then ${\merith < \meritk}$, with ${k,h~{\in \kSetPUN}}$. For the orders with the same price, the merit order is assigned according to a set of non-discriminatory rules, as for example the time stamp of submission.

Traditionally, pumping units belonging to hydroelectric production plants are excluded from the UPP rule.  These units buy electricity to refill their reservoirs usually during the night, whereas they generate energy during the remaining hours. To harmonize the buying and selling price, demand orders from these units pay zonal prices, and not the UPP. As a consequence, the set $\kSetNonPUN$ of consumers paying zonal prices is usually non-empty in the set of UPP zones $\zPUN$, because it is populated by the pumping units. By contrast, the set $\kSetPUN$ of consumers paying the UPP is always empty in the zones $\zNonPUN$ non-enforcing the UPP.

Currently, the implementation of the UPP pricing scheme on the Italian market allows an error tolerance $\imbalance \in [-1;5]$.
Therefore, the Italian PUN is actually defined as:
\begin{equation}
\pi_t \sum_{\kIndex \in \kSetPUN} \dkpi = \sum_{\zI \in \zPUN}  \sum_{\kIndex \in \kSetPUNi} \pZi\dkpi + \imbalance\label{PUNwithImbalance}\,.
\end{equation}

\subsection{Block orders}\label{blockDescription}

\begin{figure}
\centering
\includegraphics[width=0.8\linewidth]{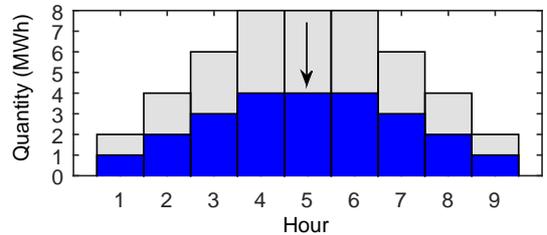}
\caption{The figure shows a curtailable profile block order (gray + blue area), which is partially executed (blue area), with acceptance ratio $\rblockAccepted=0.50$. }
\label{figProfileBlock}
\end{figure}

A block order $p$ submitted by a producer is an order that spans over multiple hours, and can be either fully accepted or fully rejected \cite{biskas2014european}. Moreover, the block order submitted price $\ppb$ must be the same over the whole timespan.
The most general form of a single block order is the \textit{profile} block order, which allows to submit different quantities for each hour. Furthermore, an additional feature called \textit{minimum acceptance ratio} (MAR) has been introduced in the Nordic countries, which allows to partially execute a block order \cite{NordPoosite}.  In this case, the hourly quantities involved can be partially cleared, and the profile block order is uniformly scaled over the whole timespan, as depicted in Figure \ref{figProfileBlock}. The fraction of the quantity executed  for each hour is termed the acceptance ratio $\rblockAccepted$. The acceptance ratio is  independent of $t$, and satisfies the constraint, $\mar \leq \rblockAccepted \leq 1$, where $\mar$ is the MAR for the block order $p$. For this reason, block orders with MAR are called \textit{curtailable}.

Block orders can be classified according to their \textit{degree of moneyness}, that is, a block order submitted by a producer is termed:
\begin{itemize}
\item in-the-money (ITM), if the submitted price $\ppb$ is smaller than the average of the zonal prices $\pZi$ weighted by the hourly offered quantities $\spbmax$, or equivalently, if the block order has a strictly positive surplus, i.e., $\sum_{t \in \tSetBlockp} \spbmax (\pZi - \ppb) > 0$;
\item at-the-money (ATM), if the submitted price $\ppb$ is equal to the average of the zonal prices $\pZi$ weighted by the hourly offered quantities $\spbmax$, or equivalently, if the block order has a zero surplus, i.e., $\sum_{t \in \tSetBlockp} \spbmax (\pZi - \ppb) = 0$;
\item out-of-the-money (OTM), if the submitted price $\ppb$ is greater than the average of the zonal prices $\pZi$ weighted by the hourly offered quantities $\spbmax$, or equivalently, if the block order has a strictly negative surplus, i.e., $\sum_{t \in \tSetBlockp} \spbmax (\pZi - \ppb) < 0$;
\end{itemize}
where ${\tSetBlockp \subseteq \tSet}$ is the timespan of block order $p$.
\textcolor{\mycolor}{We recall that the acceptance ratio $\rblockAccepted$ does not depend on time. For this reason, in the definitions of the block order surplus only $\spbmax$ is considered instead of $\rblockAccepted \spbmax$.}
ITM block orders should be fully executed. ATM block orders can be partially cleared.  OTM block orders must be always rejected. Notice that, the indivisible nature of regular block orders may prevent the existence of an optimal market equilibrium, i.e., the clearing problem may result infeasible \cite{vanVyveMadani2014minimizing,vanVyveMadani2016NonConvexities}. For this reason, current market rules allow to reject ITM block orders \cite{EUPHEMIAmanual}. Rejected ITM block orders are termed paradoxically rejected block orders (PRBs). By contrast, OTM block orders that are accepted are termed paradoxically accepted block orders (PABs). PABs are not allowed in the European markets. The reasoning behind this different treatment is straightforward. Under perfect competition, prices submitted by producers are the marginal costs \cite{kirschen2004fundamentals,rubinfeld2013microeconomics}. Therefore, a PAB would cause a monetary loss to the producer, whereas a PRB leads to a missed trading opportunity. Notice that, in some US markets \cite{uplift_in_RTO_ISO} and in Turkey \cite{turkey_day_ahead}, it is possible to compensate producers with side payments \cite{HUPPMANN2018622}, but this is not allowed in the European markets.

\subsection{Literature review}

In the literature, algorithms to solve the market clearing problem in the presence of block orders are based on different techniques. Reference \cite{biskas2016europeanElectricPowerSystemsResearch} formulates a mixed-integer linear program (MILP) to clear the market with an additional iterative process to handle PRBs. Reference \cite{vanVyveMadani2015relaxationBender} proposes a primal-dual formulation of the market clearing problem, where an improved Benders-like decomposition method is further introduced to strengthen the classical Benders cuts, which is extended in \cite{vanVyveMadani2017blockmic}. In \cite{vanVyveMadani2014minimizing} a clearing method to minimize the impact of PRBs on the final solution is proposed. Reference \cite{zakBlock} introduces a bilevel approach to handle regular block orders in a single-zone market where block order surpluses are explicitly considered. The official European algorithm for market coupling, termed EUPHEMIA \cite{EUPHEMIAmanual}, is based on a mixed-integer quadratic programming formulation with additional sequential subproblems and modules. It is partially derived from the COSMOS \cite{COSMOS} model, originally employed in the central-western European electricity markets. Both algorithms implement a branch-and-bound method for solving a European social welfare maximization problem, where appropriate cuts are introduced until an optimal solution fulfilling all the market requirements is achieved. Reference \cite{meeus2009BlockOrderRestrictions} reports an interesting scenario analysis, which investigates the effects of different size, number and type of block orders on the computation time required to solve the market clearing problem. \textcolor{\mycolor}{Reference \cite{Bovo2018} formulates a mixed-integer quadratic program to mimic the complete European day-ahead market, with iterative processes to handle PRBs, complex orders, and the UPP.}

With respect to the UPP scheme, reference \cite{biskasvlachos2011balancing} proposes a complementarity approach to solve a clearing problem under mixed pricing rules, which is further extended to reserves in \cite{biskasvlachos2011simultaneous}, and block orders in \cite{biskaschatzigiannisPUN}. In the latter case, an iterative process is implemented which involves an initial MILP problem to handle block orders, followed by a mixed-complementarity problem to deal with the UPP. \textcolor{\mycolor}{Reference \cite{biskas2015EuropeanMarketCouplingAlgorithm} uses a complementarity approach to clear a market with block and complex orders, which is extended to UPP in \cite{Dourbois2017PowerTech}, where in both cases a heuristic process is used to handle paradoxical orders.} Reference \cite{iacopoTPRWS2017} proposes a bilevel model to clear a UPP market with simple stepwise orders, where the objective function maximize the surplus of the consumers. Reference \cite{sleiszRaisz2017integratedUPP} proposes an income based approach to overcome the non-linearities of the UPP, however this model cannot be extended to curtailable block orders. Originally, the method to clear the Italian market was based on \cite{UPPOmanual}, whereas the current approach implemented by EUPHEMIA is described in \cite{EUPHEMIAmanual}. In the first case, the UPP is sequentially selected among each possible price in the aggregate market demand curve until the whole curve is explored. Then, the optimal solution is chosen among the feasible candidates that clear the market and yield the greatest social welfare. Similarly, EUPHEMIA explores the aggregate market demand curve until a feasible solution is found that clears the market while satisfying the UPP definition \eqref{PUNwithImbalance} within the error tolerance $\imbalance$. Notice that the UPP scheme differs substantially from both the consumers payment minimization scheme \cite{blanco2014consumer} (where the objective is to minimize the total consumers' payments), and from the clearing rule used in some US markets, such as \cite{CAISO}, where the common price paid by consumers is computed by using an ex-post iteration. \textcolor{\mycolor}{Finally, reference \cite{Bovo2017} reports an interesting analysis of the European market coupling impact on the UPP, where the positive effect of the increased market liquidity has been assessed.}

\subsection{Market clearing issues in the presence of block orders and UPP}

Block orders and the UPP rule pose a considerable burden on the European market clearing problem. In particular, block orders introduce at least two kinds of relevant issues. Firstly, the indivisible nature of these orders forces the introduction of binary variables. Secondly, block orders span over multiple trading hours, and impose multi-temporal constraints. On the other hand, under the UPP scheme it is not possible to use directly the traditional social welfare maximization method to clear the market, due to the possible difference in the price paid by consumers (the UPP) and the price received by producers (the zonal price) within the same zone. Moreover, the UPP scheme requires to clear consumers and producers simultaneously, because both have price-elastic curves, i.e., the demanded and offered quantities depend on the actual market prices. Furthermore, under the  marginal pricing framework \cite{schweppe1988spot,schweppeCaramanis1982optimal}, market prices are defined as the dual variables of the power balance constraints. This means that the problem formulation must involve the dual variables. Finally, the UPP definition \eqref{PUNdef} implies the presence of bilinear terms involving both quantities and prices, i.e., primal and dual variables, which make the problem non-linear and non-convex. As a consequence, the European market clearing problem with both block orders and the UPP can be regarded as a non-linear integer program, involving  both primal and dual variables with complementary and multi-temporal constraints.

\subsection{Paper contribution}

The problem of finding a computationally tractable and exact social welfare maximization formulation, for solving the market clearing problem in the presence of both block orders and the UPP scheme, is an important open issue in the European context.  The purpose of this paper is to present a non-iterative solution to this problem, which results in a MILP model, that can be solved with off-the-shelf solvers. This model is obtained starting from a non-linear integer bilevel problem, which is transformed into an equivalent single level model by using primal-dual relations and properties. Then, all the non-linearities are removed by using both standard integer algebra and an equivalent reformulation of the UPP definition. We remark that, this approach is homogeneous in spite of the different traded instruments and market rules. That is, the proposed framework deals with both block orders and the UPP by using the same comprehensive model under the exact European social welfare maximization objective, with no iterative process or subproblems. Furthermore, by construction, market prices are guaranteed to fulfill the marginal pricing scheme \cite{schweppe1988spot} as required by the European regulatory framework \cite{GuidelineCACM2015n1222} and coherently with standard market practices \cite{EUPHEMIAmanual,OMIEsite}. The solution is exact, with no approximation at least up to the level of resolution of current market data. Finally, the MILP formulation allows to prove the optimality of the solution. To summarize, the main novelties presented in this work are:
\begin{enumerate}
\item  the exact formulation of the market clearing problem in the presence of both curtailable profile block orders and the UPP as a non-linear integer bilevel problem, which is then transformed into an equivalent MILP model;
\item the use of complementary relations and integer methods to linearize the UPP definition;
\item the non-iterative approach maximizing the exact social welfare in the presence of both the UPP and curtailable block orders.
\end{enumerate}

The aim of this work is to show that the UPP scheme (i.e., a non-linear program), and block orders (which involve binary variables), that are currently handled heuristically, can be recast as a single, all-encompassing MILP problem, fulfilling all the European regulatory requirements. \textcolor{\mycolor}{This flexible approach allows one to gain knowledge on the overall clearing problem. By providing insights into the problem structure, the model can be used by transmission system operators, policy makers and stakeholders to evaluate the physical and economic impacts of both grid expansion plans and modifications to market policies and rules, by carrying out \textit{what-if} analysis on specific elements reflected in the problem objective and constraints. In this respect, we freely provide the open-source Python code of the proposed model, in order to bridge the gap between modeling and implementation, and to offer a ready-to-use tool to the interested user. Finally, we recall that the MILP formulation allows one to certify the optimality of the obtained solution.}

The remaining part of this paper is organized as follows. Section \ref{market_clearing_differences} highlights some of the clearing differences between the UPP scheme and a traditional market. Section \ref{model} presents a formulation of the non-linear integer bilevel model, and shows how the final MILP is built. Section \ref{implementation_details} illustrates some optional modeling features to detect market splits. Section \ref{results} describes the tests performed, and reports the numerical results. Finally, Section \ref{conclusions} outlines some conclusions. The complete MILP model is reported in \ref{appendixMILP}.

With respect to \cite{iacopoTPRWS2017}, curtailable profile block orders are now considered (both in UPP and non-UPP zones), a novel and equivalent UPP formulation is proposed, and the objective function represents the exact social welfare.

\section{Market clearing differences between the UPP scheme and a traditional European market}\label{market_clearing_differences}

This section provides a few examples to highlight some of the clearing differences between the UPP scheme and a traditional European market. For ease of reading, this section considers only stepwise orders.

\begin{figure}
\centering
\includegraphics[width=0.8\columnwidth]{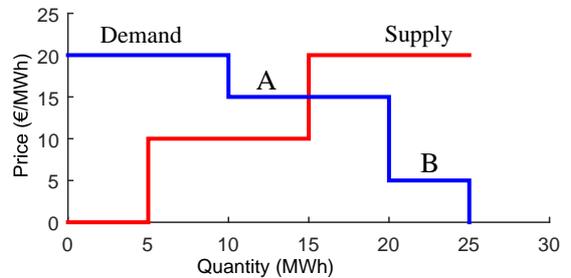}
\caption{Market clearing without the UPP rule. The intersection of the demand and supply curves determines both the zonal price and the cleared quantities. The demand order labeled by A is an at-the-money order, and it is partially cleared. By contrast, the demand order labeled by B is an  out-of-the-money order, and must be rejected.}
\label{fig1}
\end{figure}

In a traditional European market, i.e., a market cleared according to a social welfare approach with no UPP involved, the intersection of the supply and the demand curves determines both the quantity executed and the zonal price, as depicted in Figure \ref{fig1}. In this case, the demand and supply orders are cleared at the same price, i.e., the zonal price. In particular, a demand order is in-the-money if its price is strictly greater than the zonal price, whereas it is at-the-money if its price is exactly equal to the zonal price, and it is out-of-the-money if its price is strictly lower than the zonal price.
In Figure \ref{fig1}, the demand order labeled by A is intersected by the supply curve, the intersection determines the quantity partially executed, and the price of the order A sets the zonal price in the zone. \cite{kirschen2004fundamentals,rubinfeld2013microeconomics}. The order A is an at-the-money order. By contrast, the demand order labeled by B is out-of-the-money, because its price is strictly lower than the zonal price, and it must be fully rejected. This is not necessarily true for a UPP demand order, because it is cleared at the UPP and not at the zonal price.

\begin{figure}
\centering
\includegraphics[width=0.8\columnwidth]{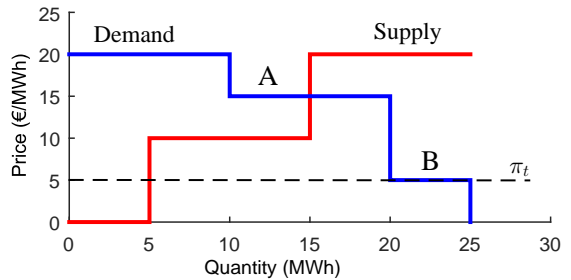}
\caption{Market clearing with the UPP rule. The UPP is assumed to be 5 Euro/MWh. The demand order labeled by A has a price of 15 Euro/MWh, whereas demand order labeled by B has a price of 5 Euro/Mwh. According to the UPP rule, the order A is in-the-money and must be fully executed. By contrast, the order B is at-the-money and can be partially cleared.}
\label{fig2}
\end{figure}
Figure \ref{fig2} shows the same demand and supply curves as in Figure \ref{fig1}. However, in this example the UPP rule is enforced, and the UPP is assumed equal to 5 Euro/MWh. Here, all the demand orders are cleared at the UPP and not at the zonal price. Therefore, the demand order labeled by A is in-the-money and must be fully executed, whereas the demand order labeled by B is at-the-money and can be partially cleared. Notice that, the order labeled by B can be partially executed regardless of the zonal price in the zone. This is a fundamental difference with respect to a traditional market.

\begin{figure}
\centering
\includegraphics[width=0.8\columnwidth]{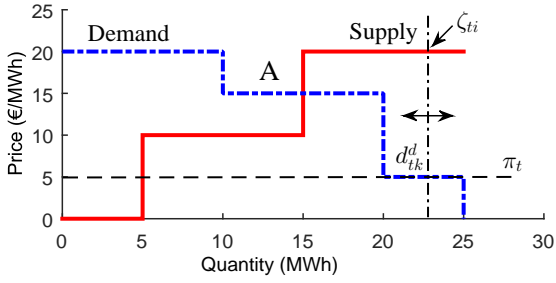}
\caption{Solution of the market clearing problem depicted in Figure \ref{fig2}. The UPP is assumed to be 5 Euro/MWh. The total demanded quantity includes the quantity $\dkPUNd$ partially cleared. The zonal price $\pZi$, collected by the producers, is determined as the price required to dispatch the demanded quantity.}
\label{fig3}
\end{figure}

In the case depicted in Figure \ref{fig2}, the in-the-money demand orders and the quantity partially cleared of the order B must be executed. Therefore, the problem boils down to finding the exact executed quantity and the zonal price, given the UPP. We recall that producers collect zonal prices. Therefore, the problem to solve is to find the price required by producers to match the demanded quantity. This problem is an economic dispatch of a potentially variable demand (which in turn depends on the UPP). In a dispatch problem, the demand is a constant, and the demand curve is considered as if it were inelastic, i.e., a vertical line, and the intersection with the supply curve determines the price required by producers, as depicted in Figure \ref{fig3}.
In this case, the demand includes the quantity partially cleared $\dkPUNd$. The zonal price $\pZi$ is the prices required by the producers to match the demanded quantity. An important consequence of the UPP pricing scheme is that it is possible to have, in the same zone, an at-the-money UPP demand order and a supply order with two different prices, both partially cleared, as in Figure \ref{fig3}. This is not possible in a traditional market as the one depicted in Figure \ref{fig1}, and it is an additional issue of the UPP pricing scheme.

\begin{figure}
\centering
\includegraphics[width=0.8\columnwidth]{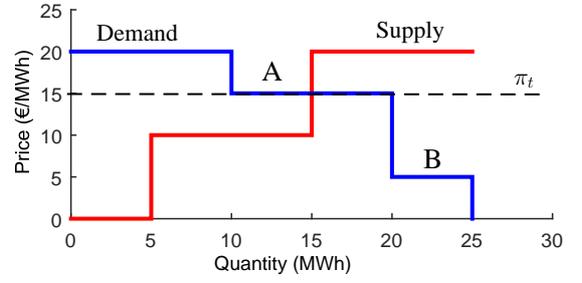}
\caption{Market clearing with the UPP rule. The UPP is assumed to be 15 Euro/MWh. This figure depicts a special case where the market equilibrium given by the intersection of the supply and demand curves, as in a traditional market, respects also the UPP rule. The demand order labeled by A is partially executed, whereas the order B must be rejected.}
\label{fig4}
\end{figure}

In particular cases, the market equilibrium given by the intersection of the demand and supply curves satisfies also the UPP rule, as shown in Figure \ref{fig4}. In this figure, the UPP is assumed to be 15 Euro/MWh. The demand order labeled by A has a price of 15 Euro/MWh, it is therefore at-the-money, and can be partially executed. This order is intersected by the supply curve. The quantity partially cleared and the zonal price determined by the intersection, as in a traditional market, fulfills also the UPP rule, because all the in-the-money UPP orders are fully executed, whereas the order partially cleared A is at-the-money. As a consequence, in this particular case, the zonal price, the UPP and the price of the demand order A coincide, i.e., $\pk=\pi_t=\pZi$. Both the cases depicted in Figure \ref{fig3}, and Figure \ref{fig4} will be considered. The first case allows a UPP demand order to be partially executed regardless of the zonal price, by relying on a dispatch approach. The second one is a special case, where we will exploit the elasticity of the demand curve and the traditional social welfare approach to deal efficiently with these at-the-money UPP orders, as shown in Section \ref{finalMILP}.

\section{The Model}\label{model}

This section presents a formalization of the market clearing problem in the presence of both the UPP and curtailable profile block orders as a non-linear integer bilevel model. Then, it shows how the bilevel model can be transformed into an equivalent MILP problem.

\subsection{Bilevel programming}

A bilevel model can be regarded as two nested optimization problems, termed upper and lower level problems \cite{bard1998practical}. Formally, a bilevel model is defined as:
\begin{align}
\max_{u \in \mathcal{U}} &\,F(u,x^*)\label{sketchUpper}\\
\text{s.t.}\,\,\,&x^* = \arg \max_{x \in \mathcal{X}} f(x;u) \, ,\label{sketchLower}
\end{align}
where $F$ and $f$ are the upper and lower level objective functions, respectively. The main feature of a bilevel program is that the upper level decision variables, labeled  $u$ in \eqref{sketchUpper}-\eqref{sketchLower}, enter the lower level as fixed parameters. The variables $x^*$ represent the optimal solution of the lower level problem, which depends on the upper level variables $u$, i.e., $x^*\,{=x^*(u)}$. However, for ease of reading this dependence is usually not formally expressed.
Historically, the bilevel approach was used in the field of game theory to describe non-cooperative Stackelberg games \cite{bard1998practical}. In a Stackelberg game the upper level problem represents a leader that acts before a follower, that is represented by the lower level problem. However, in power system economics, the bilevel method is typically used to access the dual variables, i.e., the market prices, and not to actually build a game. Therefore, the upper and the lower level objective functions,  i.e., $F$ and $f$, are usually equivalent. The interested reader is referred to \cite{bard1998practical,blanco2012unified,conejo2012complementarity} for additional information on bilevel programs and their applications in power system economics.

\subsection{The non-linear integer bilevel model}\label{bilevelFormalization}

\begin{figure}
\centering
\includegraphics[width=0.75\columnwidth]{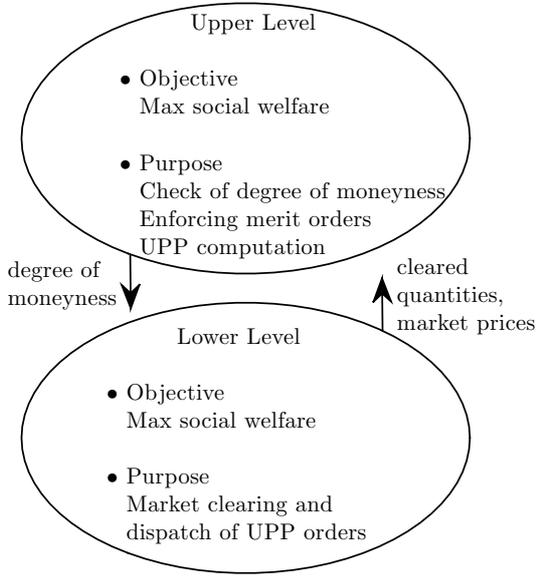}
\caption{An overview of the proposed bilevel model. The upper level handles the UPP and determines the orders' degree of moneyness. The lower level clears the market while dispatching UPP orders depending on the UPP rule.}
\label{figGraficoBilevel}
\end{figure}

In the proposed approach, the upper level problem handles the UPP and verifies the degree of moneyness of block orders, whereas the lower level actually clears the market by using a social welfare maximization approach while properly dispatching the UPP orders, as outlined in Fig.\ref{figGraficoBilevel}.

\subsubsection{The upper level problem}

This section describes the upper level problem, which is defined as follows:

\hspace{-1em}\begin{minipage}{1\columnwidth}
\begin{flalign}
\max_{\substack{\ugk, \uek, \uwk,  \\\udk, \ubp, \pi_t,\\ \dkpi, \dkPUNd, \imbalance. }}  & \sum_{t \in \tSet}  \sum_{k \in \kSetNonPUN} \pk {\dkNonPUN}^* + \sum_{t \in \tSet} \sum_{\kIndex \in \kSetPUN} \pk \dkpi \nonumber\\
&- \sum_{t \in \tSet} \sum_{p \in \pSet} \pp \sp^* - \sum_{p \in \pSetBlock} \sum_{t \in \tSetBlockp}  \ppb \rblockAccepted^* \spbmax\raisetag{18pt}
 \label{OBJUpperLevel}&&
\end{flalign}
\vspace{-2em}
\begin{flalign}
&\text{subject to:}  &&\nonumber\\
& \pi_t \sum_{\kIndex \in \kSetPUN} \dkpi = \sum_{\zI \in \zPUN}  \sum_{\kIndex \in \kSetPUNi} \pZi^*\dkpi + \imbalance && \hspace{-1em}\forall t \in \tSet \label{PUNdefUpperLevel}\\
&\pk - \pi_t \leq \bigMpi \ugk &&\hspace{-1em} \forall t \in \tSet, \, \forall \kIndex \in \kSetPUN\label{ugkConstraintUpperLevel1}\\
&\pk - \pi_t \geq \epsilon - \bigMpi (1-\ugk) && \hspace{-1em}\forall t \in \tSet, \, \forall \kIndex \in \kSetPUN \label{ugkConstraintUpperLevel2}\\
&\uek(\pk - \pi_t) = 0 &&  \hspace{-1em}\forall t \in \tSet, \, \forall k \in \kSetPUN \label{uekConstraintUpperLevel}\\
&\ugh \geq \ugk  &&   \hspace{-8em} \forall t \in \tSet, \,  \forall h,\kIndex \in \kSetPUN: \merith < \meritk\label{ugkChainUpperLevel}\raisetag{12pt}\\
&\dkpi = \ugk \dkmax + {\dkPUNw}^* + \dkPUNd  && \hspace{-1em}\forall t \in \tSet, \, \forall \kIndex \in \kSetPUN \label{dkpiDefUpperLevel}\raisetag{12pt}\\
&\sum_{t \in \tSetBlockp} \spbmax (\pZi^* - \ppb)\geq -\bigMBlock(1-\ubp) && \hspace{-1em}\forall \iIndex \in \zAll, \, \forall \pIndex \in \pSetBlocki \label{surplusBloccoUpperLevel} \raisetag{24pt}\\
&\uwk + \udk \leq \uek           &&  \hspace{-1em}\forall t \in \tSet, \, \forall k \in \kSetPUN \label{udkuwkConstraint}\\
&\dkPUNd \leq \dkmax \udk &&  \hspace{-1em}\forall t \in \tSet, \, \forall k \in \kSetPUN \label{dkPUNdMax}\\
&\ugk \in \{0,1\} \, , \, \uek \in \{0,1\} \, && \hspace{-1em}\forall t \in \tSet, \,  \forall k \in \kSetPUN \raisetag{12pt}\\
&\uwk \in \{0,1\} \, , \, \udk \in \{0,1\} \, && \hspace{-1em}\forall t \in \tSet, \,  \forall k \in \kSetPUN \raisetag{12pt}\\
&\ubp \in \{0,1\} && \hspace{-1em}\forall \pIndex \in \pSetBlock \, , \label{ubpDelUpperLevel}
\end{flalign}
\vspace{0pt}
\end{minipage}
with  $\dkpi \geq 0$, $\dkPUNd \geq 0$, $\imbalance \in [-1;5]$, and $\pi_t \in \mathbb{R}$. The term $\epsilon$ is a sufficiently small positive parameter, whereas $\bigMpi$, and $\bigMBlock$ are appropriate large constants, and a discussion on the selection of these parameters is given in \ref{appendixMILP}. Notice that, the starred variables ${\dkNonPUN}^*$, ${\dkPUNw}^*$, $\sp^*$, $\pZi^*$, and $\rblockAccepted^*$, are the optimal values of the lower level variables, as sketched in \eqref{sketchUpper}-\eqref{sketchLower}. The upper level is a social welfare maximization problem where  the first two terms in the objective function \eqref{OBJUpperLevel} represent the demand orders, the third term represents producers submitting simple stepwise orders, and the last term represents producers submitting curtailable profile block orders. Constraint \eqref{PUNdefUpperLevel} is the UPP definition stated in \eqref{PUNwithImbalance}. Constraints \eqref{ugkConstraintUpperLevel1}-\eqref{ugkConstraintUpperLevel2} imply that the binary variable $\ugk$ is equal to one if and only if the submitted price $\pk$ is \textit{strictly} greater than the UPP $\pi_t$. Whereas, constraint \eqref{uekConstraintUpperLevel} implies that the binary variable $\uek$ can be equal to one only if the submitted price $\pk$ is \textit{exactly} equal to the UPP $\pi_t$. Notice that, $\ugk$ and $\uek$ cannot be equal to one at the same time. Constraint \eqref{ugkChainUpperLevel} enforces the priority due to the merit orders, i.e., it determines the sequential execution of the in-the-money  UPP demand orders within the UPP zones, with a significant reduction in the search space of the binary variables.
\textcolor{\mycolor}{In this constraint, $h$ and $k$ are indices representing all the consumers paying the UPP, that is $h, k \in \kSetPUN$. If the order of consumer $h$ has a smaller merit order than the order of consumer $k$ (i.e., $\merith < \meritk$), then the order of consumer $h$ must be executed before the order of consumer $k$.}
Notice that merit orders are inputs, therefore constraint \eqref{ugkChainUpperLevel} is linear. Equation \eqref{dkpiDefUpperLevel} defines the auxiliary variables $\dkpi$, which are used to recap the executed quantities for UPP demand orders into single variables. Constraint \eqref{surplusBloccoUpperLevel} verifies the degree of moneyness for block orders, and implies that the binary variables $\ubp$ can be equal to one only if the block order has a non-negative surplus. That is, if the block order is accepted, i.e., $\ubp{=1}$, then the block order must be either ITM or ATM. By contrast, a block order can be rejected, i.e., $\ubp=0$, regardless of the surplus. Therefore, this formulation excludes any PAB, i.e., an OTM block order which is accepted, but it allows PRBs, i.e., ITM block orders which are rejected, consistently with the European market requirements, as described in Section \ref{blockDescription}. The constraint \eqref{udkuwkConstraint} defines the binary variables $\uwk$ and $\udk$. These variables can differ from zero only if $\uek=1$, i.e., if the UPP demand order is at-the-money, which is the requirement for having a UPP order partially executed. The variables $\uwk$ handle the case of partial execution according to a traditional social welfare approach, as depicted in Figure \ref{fig4}, whereas the variables $\udk$ handle the case of partial execution according to an economic dispatch approach, as depicted in Figure \ref{fig3}.  The constraint \eqref{udkuwkConstraint} prevents the double execution of the same order.  Finally, the constraint \eqref{dkPUNdMax} sets the limit on the maximum dispatchable quantity $\dkPUNd$. Given the upper level decision variables $\ugk$, $\uwk$, $\dkPUNd$, and $\ubp$ the market clearing is actually performed by the lower level problem.

\subsubsection{The lower level problem}

This section describes the lower level problem, which is defined as follows:

\hspace{-1.25em}\begin{minipage}{1\columnwidth}
\begin{flalign}
&\left({\dkNonPUN}\!^*,{\dkPUNw}^*,\sp^*,\rblockAccepted^*,\Fij^*,\left[\pZi^*\right]\right) = \nonumber\\
&\arg \max_{\substack{\dkNonPUN,\dkPUNw,\sp,\\ \rblockAccepted,\Fij.}} \,
\sum_{t \in \tSet}  \sum_{k \in \kSetNonPUN} \pk \dkNonPUN + \sum_{t \in \tSet} \sum_{\kIndex \in \kSetPUN} \pk \dkPUNw  \nonumber\\
&\qquad - \sum_{t \in \tSet} \sum_{p \in \pSet} \pp \sp - \sum_{p \in \pSetBlock} \sum_{t \in \tSetBlockp}  \ppb \rblockAccepted \spbmax \label{OBJLowerLevel}  \qquad \raisetag{24pt}&
\end{flalign}
\vspace{-2em}
\begin{flalign}
&\text{subject to:}\nonumber\\
&\dkPUNw \leq \uwk \dkmax	    &&	 [\vphikPUNw \geq 0]         && \hspace{-1em} \forall t \in \tSet, \, \forall \kIndex \in \kSetPUN \label{dkmax_PUN} \\
&-\dkPUNw \leq 0	    &&	 [\vphikPUNwlo \geq 0]         && \hspace{-1em} \forall t \in \tSet, \, \forall \kIndex \in \kSetPUN \label{dkmax_PUN_lo} \\
&\dkNonPUN \leq \dkmax	    &&	 [\vphikNonPUN \geq 0]         && \hspace{-1em} \forall t \in \tSet, \, \forall \kIndex \in \kSetNonPUN \label{dkmax_NonPUN} \\
&\sp \leq \spmax 	&&	 [\vphip \geq 0]         && \hspace{-1em} \forall t \in \tSet, \, \forall \pIndex \in \pSet &\label{spmaxConstraint}\raisetag{12pt}\\
&\Fij \leq \Fijmax&&	 [\deltaMax \geq 0]     \, && \hspace{-1em} \forall t \in \tSet, \, \forall \iIndex, j \in \zAll \label{FijmaxConstraint}\raisetag{12pt}\\
&\Fij + \Fji = 0 	&&	 [\etaij \in \mathbb{R}] &&  \hspace{-1em} \forall t \in \tSet, \, \forall \iIndex, j \in \zAll \label{FijEtaConstraint}\raisetag{12pt}
\end{flalign}
\vspace{-28pt}
\begin{flalign}
&\rblockAccepted \leq \ubp  	&&	 [\vphibMax \geq 0]      && \forall \pIndex \in \pSetBlock  \label{rmaxConstraint}\\
&- \rblockAccepted  \leq - \ubp \mar 	&&	 [\vphibMin \geq 0]      && \forall \pIndex \in \pSetBlock \label{rminConstraint}\raisetag{12pt}&
\end{flalign}
\vspace{-24pt}
\begin{flalign}
&\sum_{k \in \kSetNonPUNi} \dkNonPUN + \sum_{k \in \kSetPUNi} \dkPUNw - \sum_{p \in \pSeti} \sp + \sum_{j \in \zAll} \Fij
- \sum_{p \in \pSetBlocki} \rblockAccepted \spbmax = \nonumber\\
&- \sum_{k \in \kSetPUNi} \ugk \dkmax - \sum_{k \in \kSetPUNi}  \dkPUNd  \hspace{1em} [\pZi \in \mathbb{R}]  \hspace{1em} \forall t \in \tSet, \forall \zI \in \zAll \label{powerBalanceAll} \raisetag{24pt} ,&&
\end{flalign}
\vspace{0em}
\end{minipage}
with $\dkNonPUN \geq 0$, \,$\dkPUNw \in \mathbb{R}$, \, $\sp\,{\geq 0}$, $\rblockAccepted\,{\in \mathbb{R}}$, and $\Fij\,{\in \mathbb{R}}$. Dual variables are enclosed in square brackets.
Given the upper level variables $\ugk$, $\uwk$, $\dkPUNd$, and $\ubp$, the lower level problem actually clears the market while dispatching the UPP order according to their degree of moneyness. We recall that the upper level variables enter the lower level as parameters, therefore the lower level problem is a \textit{linear} program.
Notice that, the lower level objective function \eqref{OBJLowerLevel} is equivalent to the upper level objective function \eqref{OBJUpperLevel}. Indeed, if we substitute \eqref{dkpiDefUpperLevel} into \eqref{OBJUpperLevel} and considering that the terms $\pk\ugk\dkmax$ and $\pk\dkPUNd$ are constants into the lower level problem, and that any constant term can be removed from an objective function without altering the optimal solution, we obtain the equivalent lower level objective function \eqref{OBJLowerLevel}. Therefore, given the upper level variables, the lower level clears the market according to an exact social welfare maximization problem.
Constraints \eqref{dkmax_PUN}-\eqref{spmaxConstraint} impose bounds on the demanded and offered quantities. Notice that, the constraints \eqref{dkmax_PUN_lo} explicitly sets the lower bound for the demanded quantities $\dkPUNw$. This formulation will be exploited in Section \ref{finalMILP}. Constraints \eqref{FijmaxConstraint}-\eqref{FijEtaConstraint} impose bounds on the inter-zonal flows. Constraints \eqref{rmaxConstraint}-\eqref{rminConstraint} set the MAR conditions for block orders by enforcing the relation ${\mar \leq}\,\rblockAccepted\,{\leq 1}$. The binary variables $\ubp$ are used to exclude any out-of-the-money block orders as determined by \eqref{surplusBloccoUpperLevel}. We recall that, the acceptance ratio $\rblockAccepted$ must be the same during all the hours $t \in \tSetBlockp$. This means that the day-ahead clearing problem in the presence of block orders cannot be split in independent hourly subproblems. Finally, equation \eqref{powerBalanceAll} defines the power balance constraint for each zone $i\,{\in \zAll=\zPUN\cup\zNonPUN}$. The right-hand-side of \eqref{powerBalanceAll} specifies the quantities that must be dispatched. In particular, the terms $\ugk\dkmax$ represent the in-the-money orders that must be fully executed and dispatched according to the UPP clearing rule (see Section \ref{UPP_section_Intro}). By contrast, the terms $\dkPUNd$ represent the at-the-money UPP orders partially executed and to dispatch (as in case depicted in Figure \ref{fig3}), where the quantity $\dkPUNd$ is determined by the upper level. Furthermore, notice the presence of $\dkPUNw$ in the left-hand-side of \eqref{powerBalanceAll}. The variable $\dkPUNw$ is a lower level decision variable, which determines the quantity partially cleared for an at-the-money UPP order by using a social welfare approach, as in the special cases depicted in Figure \ref{fig4}. The constraint \eqref{udkuwkConstraint} prevents the double clearing of the same order. Notice further that the set $\kSetPUNi$, i.e., the consumers paying the UPP, is empty in the zones non-enforcing the UPP, that is, $\kSetPUNi\,{=\text{\O}}$ if $i\,{\in \zNonPUN}$.

The starred variables ${\dkNonPUN}\!^*$, ${\dkPUNw}^*$, $\sp^*$, $\rblockAccepted^*$, $\Fij^*$, and $\pZi^*$ represent the optimal values of the lower level variables. The zonal prices $\pZi$ are defined as the dual variables of the power balance constraints \eqref{powerBalanceAll}, as required by the marginal pricing framework \cite{schweppe1988spot,schweppeCaramanis1982optimal}. The zonal prices are used within the upper level problem to compute the UPP in equation \eqref{PUNdefUpperLevel} and to verify the degree of moneyness for block orders in the constraint \eqref{surplusBloccoUpperLevel}. In the following section, the bilevel program is reduced to a single level optimization problem.

\subsection{The equivalent single level problem}\label{singleLevel}

In order to access the dual variables $\pZi$, i.e., the zonal prices, the bilevel model formalized in Section \ref{bilevelFormalization} is reformulated as an equivalent single level optimization problem.
The objective function of the single level problem is the same of the objective function of the upper level \eqref{OBJUpperLevel}, i.e.:
\begin{minipage}{1\columnwidth}
\begin{flalign}
\max_{\substack{\ugk, \uek, \uwk,\udk, \ubp,\\ \pi_t, \dkpi,
\dkNonPUN, \dkPUNw, \dkPUNd, \\  \sp, \imbalance, \rblockAccepted, \Fij, \pZi,\\ \vphikNonPUN, \vphikPUNw, \vphikPUNwlo, \vphip, \etaij,\\
\vphibMax,\vphibMin, \deltaMax.}} \, & \, \sum_{t \in \tSet}  \sum_{k \in \kSetNonPUN} \pk \dkNonPUN + \sum_{t \in \tSet} \sum_{\kIndex \in \kSetPUN} \pk \dkpi \nonumber
\end{flalign}
\vspace{-5em}
\begin{flalign}
\hspace{11em}&- \sum_{t \in \tSet} \sum_{p \in \pSet} \pp \sp \nonumber\\
&- \sum_{p \in \pSetBlock} \sum_{t \in \tSetBlockp}  \ppb \rblockAccepted \spbmax \, .&\label{OBJSingleLevel}
\end{flalign}
\vspace{0pt}
\end{minipage}
The single level problem is a unique optimization program, and there is no distinction between upper and lower parts. Hence, all the decision variables of both the problems are present in \eqref{OBJSingleLevel}.
Furthermore, we recall that the lower level problem is a linear program, because all the upper level variables enter the lower level as parameters. As a linear program, the lower level is equivalent to its  necessary and sufficient Karush-Kuhn-Tucker (KKT) conditions. Moreover, in a linear program, the KKT complementary slackness is equivalent to the strong duality property \cite{blanco2012unified,bradley1977applied,conejo2010decision}. As a consequence, the lower level problem can be introduced into the single level problem by adding the following constraints to the single level:
\hspace{-1em}\begin{minipage}{1\columnwidth}
\begin{flalign}
&\sum_{t \in \tSet} \sum_{k \in \kSetNonPUN} \pk \dkNonPUN + \sum_{t \in \tSet} \sum_{k \in \kSetPUN} \pk \dkPUNw \nonumber\\
&- \sum_{k \in \kSet} \sum_{p \in \pSet} \pp \sp
- \sum_{p \in \pSetBlock} \sum_{t \in \tSetBlockp}  \ppb \rblockAccepted \spbmax \nonumber \\
&=  \sum_{t \in \tSet} \sum_{\kIndex \in \kSetPUN} \uwk \vphikPUNw \dkmax + \sum_{t \in \tSet} \sum_{\kIndex \in \kSetNonPUN} \vphikNonPUN \dkmax + \sum_{t \in \tSet} \sum_{\pIndex \in \pSet} \vphip \spmax \nonumber\\
&+ \sum_{t \in \tSet} \sum_{\zI \in \zAll} \sum_{\zJ \in \zAll} \deltaMax \Fijmax
- \sum_{t \in \tSet} \sum_{\zI \in \zPUN} \pZi \sum_{\kIndex \in \kSetPUNi} (\ugk \dkmax + \dkPUNd) \nonumber\\
&+ \sum_{p \in \pSetBlock} \vphibMax \ubp
- \sum_{p \in \pSetBlock} \vphibMin \ubp \mar&
\label{strongDuality}
\end{flalign}
\vspace{-2em}
\begin{flalign}
&\vphikPUNw - \vphikPUNwlo + \pZi = \pk & \forall t \in \tSet, \, \forall \zI \in \zAll, \, \forall \kIndex \in \kSetPUNi
 \label{duale1}\\
&\vphikNonPUN + \pZi \geq \pk  & \forall t \in \tSet, \, \forall \zI \in \zAll, \, \forall \kIndex \in \kSetNonPUNi \\
&\vphip - \pZi \geq  - \pp & \forall t \in \tSet, \, \forall \zI \in \zAll, \, \forall \pIndex \in \pSeti \label{duale2}
\end{flalign}
\vspace{-28pt}
\begin{flalign}
&\deltaMax + \etaij + \etaji + \pZi = 0 && \hspace{-1.5em}\forall t \in \tSet, \,\forall \zI,\zJ \in \zAll \label{duale3}
\end{flalign}
\vspace{-2.5em}
\begin{flalign}
&\vphibMax - \vphibMin -  \sum_{t \in \tSetBlockp} \pZi \spbmax \nonumber\\
&= - \sum_{t \in \tSetBlockp} \ppb \spbmax  && \hspace{-5em}\forall \zI \in \zAll, \, \forall p \in \pSetBlocki \label{duale4}\\
&\text{\eqref{dkmax_PUN}-\eqref{powerBalanceAll}} \label{constraintsOfLowerLevel}\, ,
\end{flalign}
\vspace{0pt}
\end{minipage}
where \eqref{strongDuality} is the \textit{strong duality} property, which requires the equivalence between the objective functions' values in both the primal problem and the dual problem. Conditions \eqref{duale1}-\eqref{duale4} are the constraints of the dual problem, i.e., the \textit{dual feasibility}, whereas \eqref{constraintsOfLowerLevel} refers to the original constraints of the lower level problem, i.e., the \textit{primal feasibility}.

To summarize, the single level optimization problem, equivalent to the bilevel model presented in section \ref{bilevelFormalization}, is composed by the following three parts:
\begin{enumerate}
\item the objective function \eqref{OBJSingleLevel};
\item the constraints of the upper level \eqref{PUNdefUpperLevel}-\eqref{ubpDelUpperLevel};
\item the conditions representing the lower level problem \eqref{strongDuality}-\eqref{constraintsOfLowerLevel}.
\end{enumerate}

\subsection{The final MILP model}\label{finalMILP}

The single level optimization problem presented in Section \ref{singleLevel} is a non-linear integer program. To obtain the final equivalent MILP model, all the non-linearities must be removed. There are three kinds of non-linearities in the single level:
\begin{enumerate}
\item the products of a binary variable and a continuous bounded variable, as $\uek \pi_t$ in \eqref{uekConstraintUpperLevel};
\item the product $\pi_t \dkpi$ in the UPP definition \eqref{PUNdefUpperLevel};
\item the product $\pZi \sum_{\kIndex \in \kSetPUNi}\dkPUNd$ in the strong duality \eqref{strongDuality}, and in \eqref{PUNdefUpperLevel} due to \eqref{dkpiDefUpperLevel}.
\end{enumerate}

The non-linearities due to the product of a binary and a continuous bounded variable can be removed by using standard integer algebra. As an example, the product $ux$ of the binary variable $u$, and the continuous variable $x$ with bounds $\pm M$, can be replaced by an auxiliary continuous bounded variable $y$ defined as:
\begin{align}
&-Mu \leq y \leq +Mu\label{aux1}\\
&-M(1-u) \leq x - y \leq +M(1-u)\label{aux2} \, .
\end{align}
The \ref{appendixMILP} reports all the auxiliary variables actually used, with a discussion on the selection of the big-M's values.

To handle the UPP definition \eqref{PUNdefUpperLevel}, we propose a novel and equivalent formulation.
Firstly, by using \eqref{dkpiDefUpperLevel}, the UPP definition \eqref{PUNdefUpperLevel} can be written as:
\begin{align}\label{PUNStep1}
&\sum_{k \in \kSetPUN} \pi_t \ugk \dkmax + \sum_{k \in \kSetPUN}  \pi_t \dkPUNw + \sum_{k \in \kSetPUN}  \pi_t \dkPUNd \nonumber\\
&= \sum_{i \in \zPUN} \sum_{k \in \kSetPUNi} \pZi \ugk \dkmax + \sum_{i \in \zPUN} \sum_{k \in \kSetPUNi} \pZi \dkPUNw \nonumber\\
&+ \sum_{i \in \zPUN} \sum_{k \in \kSetPUNi} \pZi \dkPUNd + \imbalance \qquad\qquad\qquad \forall t \in \tSet \, .
\end{align}
In \eqref{PUNStep1}, the terms $\pi_t \ugk \dkmax$ and $\pZi \ugk \dkmax$ involve the products of a binary variable and a continuous variable, and can be handled as shown in \eqref{aux1}-\eqref{aux2}. Furthermore, due to \eqref{uekConstraintUpperLevel} and \eqref{udkuwkConstraint}, the terms $\dkPUNw$ and $\dkPUNd$ refers to at-the-money UPP orders, where $\pi_t=\pk$, by definition. As a consequence, the terms $\pi_t \dkPUNw$ and $\pi_t \dkPUNd$ in \eqref{PUNStep1} can be replaced by $\pk \dkPUNw$ and $\pk \dkPUNd$, respectively.

The term $\pZi\dkPUNw$ in \eqref{PUNStep1} is handled as follows. Firstly, by using \eqref{duale1} the zonal prices is recast as:
\begin{equation}
\pZi = \pk - \vphikPUNw + \vphikPUNwlo \, ,
\end{equation}
therefore, $\pZi\dkPUNw$ becomes:
\begin{equation}\label{forthTermUPP2}
\dkPUNw\pk - \dkPUNw\vphikPUNw + \dkPUNw\vphikPUNwlo \, .
\end{equation}
Then, we recall that to obtain the single level problem in Section \ref{singleLevel}, the lower level has been recast by using a set of equivalent necessary and sufficient conditions, and to avoid the KKT complementary slackness, the strong duality property has been used. This is desirable because the KKT complementary slackness would introduce further non-linearities. However, the strong duality guarantees that all the KKT complementary slackness conditions hold \cite{blanco2014revenue,blanco2017solution}. Therefore, we can use any subset of them for our purpose. In particular, the complementary slackness associated to the constraints \eqref{dkmax_PUN}-\eqref{dkmax_PUN_lo} are defined as follows:
\begin{flalign}
&(\dkPUNw - \uwk\dkmax)\vphikPUNw=0 \iff \dkPUNw\vphikPUNw = \uwk\dkmax\vphikPUNw\label{complementarySlacknessdk1}\raisetag{12pt}&\\
&-\dkPUNw\vphikPUNwlo=0 \, . \label{complementarySlacknessdk2}
\end{flalign}
Hence, by using \eqref{forthTermUPP2}-\eqref{complementarySlacknessdk2} the following relation can be obtained:
\begin{equation}\label{forthTermUPP3}
\pZi \dkPUNw = \dkPUNw\pk - \uwk\dkmax\vphikPUNw \, ,
\end{equation}
which involves only the product of a binary and a continuous variable, and can be handled as showed in \eqref{aux1}-\eqref{aux2}.

Finally, the only remaining non-linearity to handle is the term $\pZi\sum_{k \in \kSetPUNi}\dkPUNd$ which is present not only in \eqref{PUNStep1} but also in \eqref{strongDuality}. To deal with this term we use a binary expansion. The basic idea of a binary expansion is to convert an integer number in binary form by using binary variables \cite{binaryGupteConvexHull,pereira2005strategic,binarizeRoy2007}. In particular, the binary expansion is utilized to convert the quantity $\sum_{k \in \kSetPUNi}\dkPUNd$ in binary form, as follows:
\begin{flalign}
&\sum_{j=0}^{J} \bji 2^j = 10^{c} \sum_{\kIndex \in \kSetPUNi} \dkPUNd&& \hspace{-1em}\forall t \in \tSet, \, \forall \iIndex \in \zPUN \label{binaryExpansionUpperLevel} \, ,
\end{flalign}
where $\bji \in \{0,1\}$, $J$ is an appropriate parameter depending on the market data, and $c$ is the number of decimal digits allowed in the considered market. In the Italian market ${c\,{=3}}$. The right-hand-side in \eqref{binaryExpansionUpperLevel} represents the term under conversion, where the value $\sum_{\kIndex \in \kSetPUNi} \dkPUNd$ is multiplied by $10^c$ in order to obtain an integer number. Then, the left-hand-side actually performs the conversion in binary form. Given the value of $c$, which depends on the market specifications, the discretization performed in \eqref{binaryExpansionUpperLevel} is exact. As a consequence the following relation holds:
\begin{equation}\label{pziSumDkPUNd}
\pZi\sum_{k \in \kSetPUNi}\dkPUNd = 10^{-c} \pZi \sum_{j=0}^{J} \bji 2^j \, .
\end{equation}

Therefore, substituting the terms described, and simplifying the common terms, the UPP definition \eqref{PUNStep1}, can be equivalently recast as:
\begin{flalign}\label{PUNdefFinal}
&\sum_{k \in \kSetPUN} \pi_t \ugk \dkmax  + \sum_{k \in \kSetPUN} \pk \dkPUNd \nonumber\\
&= \sum_{i \in \zPUN} \sum_{k \in \kSetPUNi} \pZi \ugk \dkmax - \sum_{k \in \kSetPUN} \uwk \dkmax \vphikPUNw \nonumber\\
&+ 10^{-c} \sum_{i \in \zPUN} \pZi \sum_{j=0}^J \bji 2^j + \imbalance \hspace{5em} \forall t \in \tSet \, ,
\end{flalign}
which involves only the products of a binary and a continuous variable that can be handled as showed in \eqref{aux1}-\eqref{aux2}. The definition \eqref{PUNdefFinal} is exact, with no approximation provided that the parameter $c$, introduced in \eqref{binaryExpansionUpperLevel}, is selected properly.

Starting from the single level model described in Section \ref{singleLevel}, by using \eqref{PUNdefFinal} in place of \eqref{PUNdefUpperLevel}, substituting \eqref{pziSumDkPUNd} in \eqref{strongDuality}, and after removing all the non-linearities due to the products of a binary and a continuous variable as outlined in \eqref{aux1}-\eqref{aux2}, we obtain the final MILP model reported in \ref{appendixMILP}. The MILP model solves the market clearing problem in the presence of both the UPP and curtailable profile block orders by using an exact social welfare maximization approach without any heuristic or iterative methods.

\section{Implementation details}\label{implementation_details}

Under the UPP pricing scheme, all the UPP orders have a merit order, i.e., a parameter that determines a strict total ordering among the orders, as described in Section \ref{UPP_section_Intro}. All the UPP orders must be executed sequentially, according to the priority established by the merit order. However, the current implementation of the UPP pricing scheme in the European market, strictly enforces the merit order only for the in-the-money UPP orders, see \eqref{ugkChainUpperLevel}. By contrast, the merit order for the ATM UPP orders is enforced only as long as there is enough transmission capacity between the zones involved, i.e., if there is no market split. For this reason, the merit order for ATM orders is currently enforced ex-post, given the market solution. In this section, we propose a set of constraints to detect whether the connecting lines are congested or not. In particular, we introduce a set of conditions to detect market splits, and to enforce the merit order for the ATM UPP orders directly within the optimization problem. These constraints are upper level constraints. However, they are not strictly required by the current European market rules. For this reason, they are described in this section and not in Section \ref{bilevelFormalization}.
Notice that, these constraints allow one to fix a significant part of the ATM quantities  actually executed, with a significant reduction of the search space of the binary expansion \eqref{binaryExpansionUpperLevel}. Specifically, for all $t \in \tSet$, $i \in \zPUN$, and $h,k \in \kSetPUNi$, such that $\merith < \meritk$, and $\ph = \pk$, the following constraint is added:
\begin{flalign}
&d_{th}^w + d_{th}^d \geq D_{th}^{max} \uek \label{meritIntraZone} \, .&
\end{flalign}
Furthermore, for all $t \in \tSet$, $i,j \in \zPUN$, $h \in \kSetPUNi$, and $k \in \mathcal{K}^\pi_{tj}$, with $i \ne j$, such that $\merith < \meritk$, $\ph = \pk$, and $\Fijmax > 0$, the following constraints are enforced:
\begin{flalign}
& \Fij \leq \Fijmax - \epsilon^f +  \uf      \label{ufDef_1}\\
& \Fij \geq \Fijmax -\bigMFlow (1-\uf)     \label{ufDef_2}\\
& \uf \in \{0,1\}						   \label{ufDef_3}\\
& d_{th}^w + d_{th}^d \geq D_{th}^{max} \uek - \bigMDmax\uf - \bigMDmax u^{f}_{tji} \, . \label{meritInterZoneZero}&
\end{flalign}
Notice that merit orders, prices and maximum flow capacities  are inputs, therefore the above constraints are linear. The term $\epsilon^f$ is a sufficiently small parameter, whereas $\bigMFlow$ and $\bigMDmax$ are appropriate large constants, that are defined in \ref{appendixMILP}. Constraint \eqref{meritIntraZone} enforces the merit order for ATM orders within the same zone. Furthermore, from \eqref{FijmaxConstraint} and \eqref{ufDef_1}-\eqref{ufDef_3},
$\uf=1$ if and only if $\Fij = \Fijmax$, i.e., the line is congested. Constraint \eqref{meritInterZoneZero} enforces the merit order for ATM orders in zones directly connected. If the line is not saturated, then $\uf=0$, and the constraint is enforced, otherwise there is market split and the constraint is deactivated. Constraint \eqref{meritInterZoneZero} can be further generalized to zones connected through a path involving multiple lines. In this case, to enforce the constraint, all the variables $\uf$ must be zero, i.e., the lines connecting the zones $i$ and $j$ must not be saturated, otherwise there is a market split along the path, and the merit order must not be enforced. As a remark, notice that a loose formulation of $\uf$ can be implemented by using:
\begin{equation}
\uf \leq \dfrac{\Fij + F^{max}_{tji}}{\Fijmax + F^{max}_{tji}}
\end{equation}
instead of \eqref{ufDef_1}-\eqref{ufDef_2}. In this case if $\Fij < \Fijmax$, then ${\uf{=0}}$, but the converse is not true. This approach appears to be preferable for dealing with solvers or modeling languages that do not allow to prioritize the binary variables. Indeed, the merit order for ATM orders can be regarded as a secondary requirement, and it should be enforced by the solver only at the end of the Branch-and-Bound process. This can be performed efficiently by setting the lowest priority to the binary variables $\uf$.

\section{Numerical Results}\label{results}

This section describes the numerical results obtained by testing the MILP model introduced in Section \ref{finalMILP}, and reported in \ref{appendixMILP}.

\subsection{Experiment setup}\label{dataDescription}

The data used to test the proposed MILP model was downloaded from the website of the Italian market operator \cite{sitoGME}. It refers to the day-ahead market, and covers 31 days, ranging from January $1^{st}$ to January $31^{st}$, 2018. Each day involves on average 20,307 demand orders and 37,810 offer orders. These orders are distributed over 22 zones. Six zones are the Italian physical zones, which enforce the UPP scheme, whereas the remaining zones do not enforce the UPP. We recall that the Italian UPP is termed PUN. Artificial orders were randomly added to test curtailable profile block orders. Specifically, for a block order $p$, the maximum hourly quantity offered $\spbmax$ was sampled from a uniform distribution ranging from $1$ MWh to $75$ MWh, whereas the block order price $\ppb$ was generated by using a normal distribution with mean $50$ \euro/MWh and standard deviation of $10$ \euro/MWh. The MAR is set to 10\%. The MILP model was implemented in Python 2.7 with Pyomo 5.2 \cite{pyomoBook}, and solved with CPLEX 12.5 \cite{cplex2009v12} on a 8-core 2.40GHz Intel(R) Xeon(R) CPU E5-2630 v3,  with 32 GB of RAM. The Python code is freely accessible on GitHub \cite{openDAMGitHub}, and the documentation of the main functions is available at \cite{openDAMDoc}. The documentation describes also how to obtain the Italian data from \cite{sitoGME}. Orders without a price limit, listed in the Italian market with a price $\pk{=3000}$ Euro/MWh, are assumed to be fully executed. The value of the parameter $\epsilon$ in \eqref{ugkConstraintUpperLevel2} can be set arbitrarily small.  Since the Italian market limits the resolution of the PUN to six digits, any value not greater than $10^{-6}$ is acceptable. The tuning of the big-M parameters in \eqref{ugkConstraintUpperLevel1}-\eqref{ugkConstraintUpperLevel2}, and \eqref{surplusBloccoUpperLevel} is discussed in \ref{appendixMILP}.

\subsection{Test with real Italian data}

The first test takes into account only the real data of the Italian day-ahead market, which ranges from January $1^{st}$ to January $31^{st}$, 2018, and includes 1,801,652 market orders. The purpose is to verify the effectiveness of the proposed approach in solving the UPP clearing scheme. The test involves 744 hourly PUN problems. On average, each instance of the problem contains $985$ binary variables, and is solved to optimality in $8.74$ seconds. In the performed tests, all the cleared quantities match the real quantities executed on the Italian market. The binary expansion \eqref{binaryExpansionUpperLevel} is actually utilized only in 9 cases out of 744 (1.21\% of the cases). That is, in these cases the solution contains at least one order with $\udk=1$.
\begin{figure}
\centering
\includegraphics[width=1\linewidth]{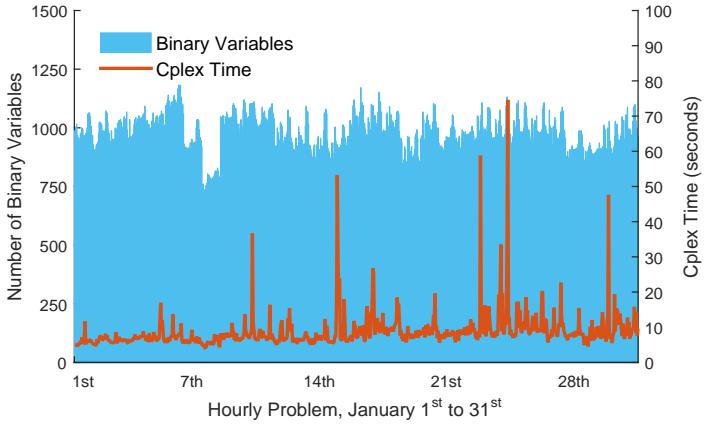}
\caption{Test of the proposed MILP model using real data of the Italian day-ahead market, from January $1^{st}$ to January $31^{st}$, 2018. The test involves 744 independent hourly problems. For each instance of the problem, the figure reports the number of binary variables (bar chart, left axis), and the computation time (line chart, right axis) to reach the optimal solution.}
\label{figTempoBinarie}
\end{figure}
Figure \ref{figTempoBinarie} reports for each hourly PUN problem the number of binary variables involved, and the time to reach the optimal solution. The largest spikes in the computation time correspond to the instances where the binary expansion actually takes place. The maximum time is 74.60 seconds, which corresponds to the 20th hour of January 24th.

\subsection{Test with 50 curtailable profile block orders over 12 hours}

The second test involves the data of the Italian market operator referring to January $1^{st}$, with the addition of $50$ curtailable profile block orders randomly generated as described in Section \ref{dataDescription}. Each block order spans from the 9th to the 20th hour.  To test the effectiveness of the proposed MILP model, the block orders are evenly distributed between an PUN zone (Sicily) and a non-PUN zone (Swiss). The presence of block orders requires to solve a single MILP problem spanning the whole considered day. The clearing problem involves 19,246 PUN demand orders, 414 non-UPP demand orders, 34,927 supply orders, 50 curtailable profile block orders, and
is solved in $848.79$ seconds.

Table \ref{table1} reports the real Italian PUN (second column), and the PUN obtained from the proposed model (third column). The shadowed rows correspond to the hours where the block orders have been added. As can be observed, the difference between the real and the modeled PUN is zero up to the fourth decimal place in the hours where the block orders are not present. This discrepancy is due to the tolerance parameter $\kappa_t$ in \eqref{PUNdefUpperLevel}, which can lead to slight differences in the PUN, despite the same matched quantities.
By contrast, from the 9th to the 20th hour, the presence of block orders leads to a decrease in the PUN, which is caused by the additional quantity supplied by the block orders.

\begin{table}[h!]
\centering
\setlength{\tabcolsep}{1pt}
\setlength\extrarowheight{1pt}
\caption{Test with real Italian market data and 50 curtailable profile block orders over 12 hours}
\label{table1}
\begin{tabular}{|c|c|c|c|c|}
\hline
Hour & PUN (Real) & PUN with block orders & Difference  \\
     & (\euro/MWh)& (\euro/MWh)     & (\euro/MWh) \\ \hline
h1   & 45.822320  & 45.822277       & -0.000043   \\ \hline
h2   & 44.160000  & 44.159954       & -0.000046   \\ \hline
h3   & 42.240000  & 42.239952       & -0.000048   \\ \hline
h4   & 39.290000  & 39.289950       & -0.000050   \\ \hline
h5   & 36.000000  & 35.999949       & -0.000051   \\ \hline
h6   & 41.990000  & 41.989950       & -0.000050   \\ \hline
h7   & 42.250000  & 42.249953       & -0.000047   \\ \hline
h8   & 44.970000  & 44.969955       & -0.000045   \\ \hline
\rowcolor[HTML]{C0C0C0}
h9   & 45.000000  & 43.239955       & -1.760045   \\ \hline
\rowcolor[HTML]{C0C0C0}
h10  & 44.940000  & 43.819958       & -1.120042   \\ \hline
\rowcolor[HTML]{C0C0C0}
h11  & 45.024790  & 44.643418       & -0.381372   \\ \hline
\rowcolor[HTML]{C0C0C0}
h12  & 45.709640  & 45.709599       & -0.000041   \\ \hline
\rowcolor[HTML]{C0C0C0}
h13  & 46.700810  & 45.412933       & -1.287877   \\ \hline
\rowcolor[HTML]{C0C0C0}
h14  & 43.980140  & 43.652657       & -0.327483   \\ \hline
\rowcolor[HTML]{C0C0C0}
h15  & 44.961410  & 41.827956       & -3.133454   \\ \hline
\rowcolor[HTML]{C0C0C0}
h16  & 47.532150  & 45.170423       & -2.361727   \\ \hline
\rowcolor[HTML]{C0C0C0}
h17  & 49.906020  & 49.199960       & -0.706060   \\ \hline
\rowcolor[HTML]{C0C0C0}
h18  & 54.300000  & 52.699966       & -1.600034   \\ \hline
\rowcolor[HTML]{C0C0C0}
h19  & 51.910000  & 50.329967       & -1.580033   \\ \hline
\rowcolor[HTML]{C0C0C0}
h20  & 51.380000  & 50.069968       & -1.310032   \\ \hline
h21  & 49.200000  & 49.199967       & -0.000033   \\ \hline
h22  & 45.730000  & 45.729965       & -0.000035   \\ \hline
h23  & 44.840000  & 44.839962       & -0.000038   \\ \hline
h24  & 38.110000  & 38.109958       & -0.000042   \\ \hline
\end{tabular}
\end{table}

Table \ref{table2} reports the surplus of each block order and the acceptance ratio $\rblockAccepted$. All the block orders with a positive surplus, i.e., the ITM block orders, are fully cleared ($\rblockAccepted{=1}$). The block orders with a negative surplus, i.e., the OTM block orders, are correctly rejected ($\rblockAccepted{=0}$). Furthermore, notice that {block order 15} in the PUN zone has zero surplus (see the corresponding row in the left part of Table \ref{table2}). That is, it is an ATM block order, and it is partially cleared with $\rblockAccepted{=0.86}$.
\begin{table}[h!]
\centering
\setlength{\tabcolsep}{2pt}
\setlength\extrarowheight{1pt}
\caption{Surplus and acceptance ratio of the 50 curtailable profile block orders over 12 hours}
\label{table2}
\begin{tabular}{|c|c|r|c|c|c|r|}
\cline{1-3} \cline{5-7}
\multicolumn{3}{|c|}{PUN zone} & \multicolumn{1}{l|}{} & \multicolumn{3}{c|}{non-PUN zone} \\ \cline{1-3} \cline{5-7}
Block order   & $\rblockAccepted$     & Surplus  &    \hspace{1em}       & Block order   & $\rblockAccepted$  & Surplus   \\ \cline{1-3} \cline{5-7}
\# 1  & 1                 & 757.71    &                       & \# 1  & 0                 & -6006.43  \\ \cline{1-3} \cline{5-7}
\# 2  & 0                 & -3189.49  &                       & \# 2  & 1                 & 1595.23   \\ \cline{1-3} \cline{5-7}
\# 3  & 1                 & 1734.14   &                       & \# 3  & 0                 & -860.09   \\ \cline{1-3} \cline{5-7}
\# 4  & 0                 & -798.90   &                       & \# 4  & 1                 & 1588.19   \\ \cline{1-3} \cline{5-7}
\# 5  & 0                 & -1240.37  &                       & \# 5  & 1                 & 3955.22   \\ \cline{1-3} \cline{5-7}
\# 6  & 0                 & -8509.20  &                       & \# 6  & 0                 & -2752.53  \\ \cline{1-3} \cline{5-7}
\# 7  & 1                 & 1103.28   &                       & \# 7  & 1                 & 2009.12   \\ \cline{1-3} \cline{5-7}
\# 8  & 1                 & 2823.18   &                       & \# 8  & 0                 & -8152.57  \\ \cline{1-3} \cline{5-7}
\# 9  & 0                 & -1227.17  &                       & \# 9  & 0                 & -1243.52  \\ \cline{1-3} \cline{5-7}
\# 10 & 0                 & -4390.73  &                       & \# 10 & 0                 & -1014.00  \\ \cline{1-3} \cline{5-7}
\# 11 & 1                 & 3670.59   &                       & \# 11 & 1                 & 375.30    \\ \cline{1-3} \cline{5-7}
\# 12 & 0                 & -4164.72  &                       & \# 12 & 0                 & -7096.47  \\ \cline{1-3} \cline{5-7}
\# 13 & 0                 & -4077.12  &                       & \# 13 & 0                 & -5852.56  \\ \cline{1-3} \cline{5-7}
\# 14 & 1                 & 2050.83   &                       & \# 14 & 0                 & -2750.48  \\ \cline{1-3} \cline{5-7}
\# 15 & 0.86              & 0.00      &                       & \# 15 & 0                 & -5519.35  \\ \cline{1-3} \cline{5-7}
\# 16 & 1                 & 9032.51   &                       & \# 16 & 0                 & -2426.21  \\ \cline{1-3} \cline{5-7}
\# 17 & 1                 & 1892.57   &                       & \# 17 & 0                 & -1250.19  \\ \cline{1-3} \cline{5-7}
\# 18 & 1                 & 130.61    &                       & \# 18 & 1                 & 3851.04   \\ \cline{1-3} \cline{5-7}
\# 19 & 1                 & 1695.72   &                       & \# 19 & 0                 & -4958.88  \\ \cline{1-3} \cline{5-7}
\# 20 & 0                 & -170.42   &                       & \# 20 & 0                 & -191.22   \\ \cline{1-3} \cline{5-7}
\# 21 & 0                 & -2345.83  &                       & \# 21 & 1                 & 805.30    \\ \cline{1-3} \cline{5-7}
\# 22 & 0                 & -4060.25  &                       & \# 22 & 0                 & -13577.37 \\ \cline{1-3} \cline{5-7}
\# 23 & 0                 & -10139.95 &                       & \# 23 & 1                 & 3728.25   \\ \cline{1-3} \cline{5-7}
\# 24 & 0                 & -1043.00  &                       & \# 24 & 0                 & -11281.57 \\ \cline{1-3} \cline{5-7}
\# 25 & 0                 & -3251.99  &                       & \# 25 & 0                 & -3678.96  \\ \cline{1-3} \cline{5-7}
\end{tabular}
\end{table}

\section{Conclusions}\label{conclusions}

The coupling of all the European electricity markets is an ongoing process which faces several difficulties. In particular, the presence of heterogeneous orders and rules, such as block orders and the uniform purchase price, rises several issues.

The proposed mixed-integer linear program allows one to solve the market clearing problem in the presence of both curtailable profile block orders and the uniform purchase price scheme. In particular, it harmonizes within a unique optimization program two classes of heterogeneous orders and rules. An exact social welfare maximization problem is formulated and solved, as required by European guidelines. The proposed approach is non-iterative, heuristic-free, and the solution is exact, with no approximation up to the level of resolution of current market data. In addition, the solution is obtained coherently with the marginal pricing framework. Finally, the mixed-integer linear program formulation allows one to prove the optimality of the obtained solution.

Ongoing work aims at introducing linked block orders, piecewise linear orders, and complex orders (as the Iberian minimum income condition) in the proposed framework. It is expected that the income condition could be modeled in a similar way as the surplus of the block order. By contrast, piecewise linear orders will pose additional issues, due to the presence of  quadratic terms in the objective function.

\section*{Acknowledgment}
The authors would like to thank the anonymous Reviewers for their valuable and constructive comments.

\appendix

\section{The final MILP model}\label{appendixMILP}

The objective function of the final MILP model is defined in \eqref{OBJSingleLevel}, whereas the problem's constraints are:
\begin{flalign}
&\sum_{k \in \kSetPUN} \yugPUNk \dkmax  + \sum_{k \in \kSetPUN} \pk \dkPUNd = \sum_{i \in \zPUN} \sum_{k \in \kSetPUNi} \yugPzk \dkmax \nonumber\\
&- \sum_{k \in \kSetPUN} \yuwvphik  \dkmax + 10^{-c} \sum_{i \in \zPUN} \sum_{j=0}^J \ybPzi 2^j + \imbalance && \forall t \in \tSet
\end{flalign}
\vspace{-1em}
\begin{flalign}
&\sum_{t \in \tSet} \sum_{k \in \kSetNonPUN} \pk \dkNonPUN + \sum_{t \in \tSet} \sum_{k \in \kSetPUN} \pk \dkPUNw \nonumber\\
&- \sum_{k \in \kSet} \sum_{p \in \pSet} \pp \sp
- \sum_{p \in \pSetBlock} \sum_{t \in \tSetBlockp}  \ppb \rblockAccepted \spbmax \nonumber \\
&=  \sum_{t \in \tSet} \sum_{\kIndex \in \kSetPUN} \yuwvphik \dkmax + \sum_{t \in \tSet} \sum_{\kIndex \in \kSetNonPUN} \vphikNonPUN \dkmax + \sum_{t \in \tSet} \sum_{\pIndex \in \pSet} \vphip \spmax \nonumber\\
&+ \sum_{t \in \tSet} \sum_{\zI \in \zAll} \sum_{\zJ \in \zAll} \deltaMax \Fijmax
- \sum_{t \in \tSet} \sum_{\zI \in \zPUN}  \sum_{\kIndex \in \kSetPUNi} \yugPzk \dkmax
\nonumber\\
&- 10^{-c} \sum_{t \in \tSet} \sum_{\zI \in \zPUN}   \sum_{j=0}^J \ybPzi 2^j \nonumber\\
&+ \sum_{p \in \pSetBlock} \yubVphibMax
- \sum_{p \in \pSetBlock} \yubVphibMin \mar
\end{flalign}
\begin{flalign}
&\text{\eqref{ugkConstraintUpperLevel1}-\eqref{ugkConstraintUpperLevel2}}\\
&\uek\pk - \yuePUNk = 0 &&  \hspace{-1em}\forall t \in \tSet, \, \forall k \in \kSetPUN\\
&\text{\eqref{ugkChainUpperLevel}-\eqref{ubpDelUpperLevel}}\\
&\text{\eqref{dkmax_PUN}-\eqref{powerBalanceAll}}\\
&\text{\eqref{duale1}-\eqref{duale4}}\\
&\text{\eqref{binaryExpansionUpperLevel}} \, ,
\end{flalign}
where the auxiliary variables are defined as follows:
\begin{flalign}
&-\bigMpi \ugk \leq \yugPUNk \leq \bigMpi \ugk                        \label{AppendixAuxiliary1}\\
&-\bigMpi (1 - \ugk) \leq \pi - \yugPUNk  \leq \bigMpi (1 - \ugk)      \\
&-\bigMpi \uek \leq \yuePUNk \leq \bigMpi \uek                        \\
&-\bigMpi (1 - \uek) \leq \pi - \yuePUNk  \leq \bigMpi (1 - \uek)      \\
&0 \leq \yuwvphik \leq \bigMpi \uwk  					  \\
&0 \leq \vphikPUNw - \yuwvphik \leq \bigMpi (1 - \uwk)   \, ,
\end{flalign}
with $t \in \tSet$, $\kIndex \in \kSetPUN$;
\begin{flalign}
&-\bigMpi \ugk \leq \yugPzk  \leq \bigMpi \ugk   \\
&-\bigMpi (1 - \ugk) \leq \pZi - \yugPzk  \leq \bigMpi (1 - \ugk)   \, ,
\end{flalign}
with $t \in \tSet$, $\zI \in \zPUN$, $\kIndex \in \kSetPUNi$;
\begin{flalign}
&-\bigMpi\bji \leq \ybPzi \leq \bigMpi \bji  			 \\
&-\bigMpi (1 - \bji) \leq \pZi - \ybPzi \leq \bigMpi (1 - \bji) \, , \label{AppendixAuxiliary2}
\end{flalign}
with $t \in \tSet$, $\iIndex \in \zPUN$, $j \in \{0,\ldots, J\}$, where $J$ is an appropriate parameter depending on market data;
\begin{flalign}
&0 \leq \yubVphibMax \leq \bigMBlock \ubp  					 \label{AppendixAuxiliaryBlock1}\\
&0 \leq \vphibMax - \yubVphibMax \leq \bigMBlock (1 - \ubp)  \\
&0 \leq \yubVphibMin \leq \bigMBlock \ubp  					  \\
&0 \leq \vphibMin - \yubVphibMin \leq \bigMBlock (1 - \ubp)    \, ,\label{AppendixAuxiliaryBlock2}
\end{flalign}
with $p \in \pSetBlock$.

In \eqref{ugkConstraintUpperLevel2} the value of $\epsilon$ is $10^{-8}$, whereas in \eqref{ufDef_1} the value of $\epsilon^f$ is $10^{-6}$.
In the Italian market, the maximum price $\bigMpi$, used in \eqref{ugkConstraintUpperLevel1}-\eqref{ugkConstraintUpperLevel2} and \eqref{AppendixAuxiliary1}-\eqref{AppendixAuxiliary2}, is $3000$ Euro/MWh.
Moreover, considering that block orders span over multiple hours, the value of $\bigMBlock$ in \eqref{surplusBloccoUpperLevel} and \eqref{AppendixAuxiliaryBlock1}-\eqref{AppendixAuxiliaryBlock2} is set to {$\bigMpi\sum_{t \in \mathcal{T}_p} S_{tp}^{B,max}$}.
In \eqref{ufDef_2} the parameter $\bigMFlow$ is defined as $\bigMFlow=\Fijmax + F^{max}_{tji}$, whereas in \eqref{meritInterZoneZero} the parameter $\bigMDmax$ is defined as $\bigMDmax=D^{max}_{th}$.

Furthermore, in order to  reduce significantly the search space of the upper level binary variables, the following constraint can be implemented:
\begin{flalign}
&\uek \leq \ugh - \ugk  && \hspace{-7.25em} \forall t \in \tSet, \,      \forall h,\kIndex \in \kSetPUN \, ,
\end{flalign}
such that $P^d_{th} > \pk$.

\section*{References}

\bibliographystyle{elsarticle-num}
\bibliography{mybib}

\begin{thebibliography}{10}
\expandafter\ifx\csname url\endcsname\relax
  \def\url#1{\texttt{#1}}\fi
\expandafter\ifx\csname urlprefix\endcsname\relax\def\urlprefix{URL }\fi
\expandafter\ifx\csname href\endcsname\relax
  \def\href#1#2{#2} \def\path#1{#1}\fi

\bibitem{ZOU201756}
P.~Zou, Q.~Chen, Y.~Yu, Q.~Xia, C.~Kang, Electricity markets evolution with the
  changing generation mix: An empirical analysis based on china 2050 high
  renewable energy penetration roadmap, Applied Energy 185 (2017) 56 -- 67.
\newblock \href
  {http://dx.doi.org/https://doi.org/10.1016/j.apenergy.2016.10.061}
  {\path{doi:https://doi.org/10.1016/j.apenergy.2016.10.061}}.

\bibitem{MARKLEHU20181290}
J.~Märkle-Huß, S.~Feuerriegel, D.~Neumann, Large-scale demand response and
  its implications for spot prices, load and policies: Insights from the
  german-austrian electricity market, Applied Energy 210 (2018) 1290 -- 1298.
\newblock \href
  {http://dx.doi.org/https://doi.org/10.1016/j.apenergy.2017.08.039}
  {\path{doi:https://doi.org/10.1016/j.apenergy.2017.08.039}}.

\bibitem{KHAN201839}
A.~S.~M. Khan, R.~A. Verzijlbergh, O.~C. Sakinci, L.~J.~D. Vries, How do demand
  response and electrical energy storage affect (the need for) a capacity
  market?, Applied Energy 214 (2018) 39 -- 62.
\newblock \href
  {http://dx.doi.org/https://doi.org/10.1016/j.apenergy.2018.01.057}
  {\path{doi:https://doi.org/10.1016/j.apenergy.2018.01.057}}.

\bibitem{IYCHETTIRA2017228}
K.~K. Iychettira, R.~A. Hakvoort, P.~Linares, R.~de~Jeu, Towards a
  comprehensive policy for electricity from renewable energy: Designing for
  social welfare, Applied Energy 187 (2017) 228 -- 242.
\newblock \href
  {http://dx.doi.org/https://doi.org/10.1016/j.apenergy.2016.11.035}
  {\path{doi:https://doi.org/10.1016/j.apenergy.2016.11.035}}.

\bibitem{OLIVELLAROSELL2018881}
P.~Olivella-Rosell, E.~Bullich-Massagué, M.~Aragüés-Peñalba, A.~Sumper,
  S.~Ødegaard Ottesen, J.-A. Vidal-Clos, R.~Villafáfila-Robles, Optimization
  problem for meeting distribution system operator requests in local
  flexibility markets with distributed energy resources, Applied Energy 210
  (2018) 881 -- 895.
\newblock \href
  {http://dx.doi.org/https://doi.org/10.1016/j.apenergy.2017.08.136}
  {\path{doi:https://doi.org/10.1016/j.apenergy.2017.08.136}}.

\bibitem{AYON20171}
X.~Ayón, J.~Gruber, B.~Hayes, J.~Usaola, M.~Prodanović, An optimal day-ahead
  load scheduling approach based on the flexibility of aggregate demands,
  Applied Energy 198 (2017) 1 -- 11.
\newblock \href
  {http://dx.doi.org/https://doi.org/10.1016/j.apenergy.2017.04.038}
  {\path{doi:https://doi.org/10.1016/j.apenergy.2017.04.038}}.

\bibitem{DUENAS2018443}
P.~Duenas, A.~Ramos, K.~Tapia-Ahumada, L.~Olmos, M.~Rivier, J.-I.
  Pérez-Arriaga, Security of supply in a carbon-free electric power system:
  The case of iceland, Applied Energy 212 (2018) 443 -- 454.
\newblock \href
  {http://dx.doi.org/https://doi.org/10.1016/j.apenergy.2017.12.028}
  {\path{doi:https://doi.org/10.1016/j.apenergy.2017.12.028}}.

\bibitem{Forte2017}
B.~Aluisio, M.~Dicorato, G.~Forte, M.~Trovato, A.~Sallati, C.~Gadaleta,
  C.~Vergine, F.~Ciasca, The application of a flow-based methodology for yearly
  network analysis according to market data, in: 2017 14th International
  Conference on the European Energy Market (EEM), 2017, pp. 1--6.
\newblock \href {http://dx.doi.org/10.1109/EEM.2017.7981918}
  {\path{doi:10.1109/EEM.2017.7981918}}.

\bibitem{glachant2016mapping}
J.-M. Glachant, Mapping the course of the {EU} `{P}ower {T}arget {M}odel' on
  its own terms, Robert Schuman Centre for Advanced Studies Research Paper No.
  RSCAS 23.

\bibitem{strbac016benefitsofintegrating}
D.~Newbery, G.~Strbac, I.~Viehoff, The benefits of integrating {E}uropean
  electricity markets, Energy Policy 94 (2016) 253--263.

\bibitem{interactionBetweenMarkets2017}
T.~Brijs, C.~D. Jonghe, B.~F. Hobbs, R.~Belmans, Interactions between the
  design of short-term electricity markets in the {CWE} region and power system
  flexibility, Applied Energy 195 (2017) 36 -- 51.
\newblock \href {http://dx.doi.org/doi.org/10.1016/j.apenergy.2017.03.026}
  {\path{doi:doi.org/10.1016/j.apenergy.2017.03.026}}.

\bibitem{balancingMarketNorth}
H.~Farahmand, G.~Doorman, Balancing market integration in the {N}orthern
  {E}uropean continent, Applied Energy 96~(Supplement C) (2012) 316 -- 326.
\newblock \href {http://dx.doi.org/doi.org/10.1016/j.apenergy.2011.11.041}
  {\path{doi:doi.org/10.1016/j.apenergy.2011.11.041}}.

\bibitem{EUPHEMIAmanual}
{PCR PXs},
  \href{www.mercatoelettrico.org/en/MenuBiblioteca/Documenti/20160127EuphemiaPublicDescription.pdf}{EUPHEMIA
  Public Description PCR Market Coupling Algorithm}, {V}ersion 1.5.
\newline\urlprefix\url{www.mercatoelettrico.org/en/MenuBiblioteca/Documenti/20160127EuphemiaPublicDescription.pdf}

\bibitem{OMIEsite}
{Polo espa\~nol S.A.},
  \href{www.omie.es/files/reglas_20140127_ingles_no_oficial_v1.pdf}{{D}aily and
  intraday electricity market operating rules}.
\newline\urlprefix\url{www.omie.es/files/reglas_20140127_ingles_no_oficial_v1.pdf}

\bibitem{sitoGME}
{Gestore dei Mercati Energetici S.p.A.}
\newblock \href{www.mercatoelettrico.org/en/Esiti/MGP/EsitiMGP.aspx}{[link]}.
\newline\urlprefix\url{www.mercatoelettrico.org/en/Esiti/MGP/EsitiMGP.aspx}

\bibitem{EPEXsite}
{EPEX SPOT SE}.
\newblock \href{www.epexspot.com/en/product-info/auction}{[link]}.
\newline\urlprefix\url{www.epexspot.com/en/product-info/auction}

\bibitem{NordPoosite}
{Nord Pool AS}.
\newblock
  \href{www.nordpoolspot.com/TAS/Rulebook-for-the-Physical-Markets/}{[link]}.
\newline\urlprefix\url{www.nordpoolspot.com/TAS/Rulebook-for-the-Physical-Markets/}

\bibitem{biskas2014european}
P.~N. Biskas, D.~I. Chatzigiannis, A.~G. Bakirtzis, European electricity market
  integration with mixed market designs. {P}art {I}: Formulation, {IEEE} Trans.
  Power Syst. 29~(1) (2014) 458--465.

\bibitem{vanVyveMadani2014minimizing}
M.~Madani, M.~Van~Vyve, Minimizing opportunity costs of paradoxically rejected
  block orders in {E}uropean day-ahead electricity markets, in: 11th
  International Conference on the {E}uropean Energy Market (EEM), 2014, IEEE,
  2014, pp. 1--6.

\bibitem{vanVyveMadani2016NonConvexities}
M.~Madani, M.~V. Vyve, A.~Marien, M.~Maenhoudt, P.~Luickx, A.~Tirez,
  Non-convexities in {E}uropean day-ahead electricity markets: {B}elgium as a
  case study, in: 13th International Conference on the {E}uropean Energy Market
  (EEM), 2016, 2016, pp. 1--5.
\newblock \href {http://dx.doi.org/10.1109/EEM.2016.7521190}
  {\path{doi:10.1109/EEM.2016.7521190}}.

\bibitem{kirschen2004fundamentals}
D.~S. Kirschen, G.~Strbac, Fundamentals of power system economics, John Wiley
  \& Sons, 2004.

\bibitem{rubinfeld2013microeconomics}
D.~Rubinfeld, R.~Pindyck, Microeconomics, Pearson, 2013.

\bibitem{uplift_in_RTO_ISO}
F.~E. R.~C. Staff,
  \href{https://www.ferc.gov/legal/staff-reports/2014/08-13-14-uplift.pdf}{Uplift
  in {RTO} and {ISO} markets} (2014).
\newline\urlprefix\url{https://www.ferc.gov/legal/staff-reports/2014/08-13-14-uplift.pdf}

\bibitem{turkey_day_ahead}
{Energy Exchange Instambul},
  \href{https://www.epias.com.tr/wp-content/uploads/2016/12/public_document_eng_v3.pdf}{Day-ahead
  electricity market clearing algorithm}.
\newline\urlprefix\url{https://www.epias.com.tr/wp-content/uploads/2016/12/public_document_eng_v3.pdf}

\bibitem{HUPPMANN2018622}
D.~Huppmann, S.~Siddiqui, An exact solution method for binary equilibrium
  problems with compensation and the power market uplift problem, European
  Journal of Operational Research 266~(2) (2018) 622 -- 638.
\newblock \href {http://dx.doi.org/https://doi.org/10.1016/j.ejor.2017.09.032}
  {\path{doi:https://doi.org/10.1016/j.ejor.2017.09.032}}.

\bibitem{biskas2016europeanElectricPowerSystemsResearch}
D.~I. Chatzigiannis, G.~A. Dourbois, P.~N. Biskas, A.~G. Bakirtzis, European
  day-ahead electricity market clearing model, Electric Power Systems Research
  140 (2016) 225--239.

\bibitem{vanVyveMadani2015relaxationBender}
M.~Madani, M.~Van~Vyve, Computationally efficient {MIP} formulation and
  algorithms for {E}uropean day-ahead electricity market auctions, European
  Journal of Operational Research 242~(2) (2015) 580--593.

\bibitem{vanVyveMadani2017blockmic}
M.~Madani, M.~V. Vyve, Revisiting minimum profit conditions in uniform price
  day-ahead electricity auctions, European Journal of Operational Research\href
  {http://dx.doi.org/https://doi.org/10.1016/j.ejor.2017.10.024}
  {\path{doi:https://doi.org/10.1016/j.ejor.2017.10.024}}.

\bibitem{zakBlock}
E.~J. Zak, S.~Ammari, K.~W. Cheung, Modeling price-based decisions in advanced
  electricity markets, in: 2012 9th International Conference on the {E}uropean
  Energy Market, 2012, pp. 1--6.
\newblock \href {http://dx.doi.org/10.1109/EEM.2012.6254813}
  {\path{doi:10.1109/EEM.2012.6254813}}.

\bibitem{COSMOS}
{APX, BELPEX and EPEX SPOT},
  \href{http://static.epexspot.com/document/20015/COSMOS_public_description.pdf}{COSMOS
  description - CWE Market Coupling algorithm}, {V}ersion 1.1.
\newline\urlprefix\url{http://static.epexspot.com/document/20015/COSMOS_public_description.pdf}

\bibitem{meeus2009BlockOrderRestrictions}
L.~Meeus, K.~Verhaegen, R.~Belmans, Block order restrictions in combinatorial
  electric energy auctions, European Journal of Operational Research 196~(3)
  (2009) 1202 -- 1206.
\newblock \href {http://dx.doi.org/https://doi.org/10.1016/j.ejor.2008.04.031}
  {\path{doi:https://doi.org/10.1016/j.ejor.2008.04.031}}.

\bibitem{Bovo2018}
L.~H. Lam, V.~Ilea, C.~Bovo, European day-ahead electricity market coupling:
  Discussion, modeling, and case study, Electric Power Systems Research 155
  (2018) 80 -- 92.
\newblock \href {http://dx.doi.org/https://doi.org/10.1016/j.epsr.2017.10.003}
  {\path{doi:https://doi.org/10.1016/j.epsr.2017.10.003}}.

\bibitem{biskasvlachos2011balancing}
A.~G. Vlachos, P.~N. Biskas, Balancing supply and demand under mixed pricing
  rules in multi-area electricity markets, {IEEE} Trans. Power Syst. 26~(3)
  (2011) 1444--1453.

\bibitem{biskasvlachos2011simultaneous}
A.~G. Vlachos, P.~N. Biskas, Simultaneous clearing of energy and reserves in
  multi-area markets under mixed pricing rules, {IEEE} Trans. Power Syst.
  26~(4) (2011) 2460--2471.

\bibitem{biskaschatzigiannisPUN}
D.~I. Chatzigiannis, P.~N. Biskas, G.~A. Dourbois, European-type electricity
  market clearing model incorporating {PUN} orders, {IEEE} Trans. Power Syst.
  32~(1) (2017) 261--273.
\newblock \href {http://dx.doi.org/10.1109/TPWRS.2016.2542823}
  {\path{doi:10.1109/TPWRS.2016.2542823}}.

\bibitem{biskas2015EuropeanMarketCouplingAlgorithm}
G.~A. Dourbois, P.~N. Biskas, European market coupling algorithm incorporating
  clearing conditions of block and complex orders, in: PowerTech, 2015 IEEE
  Eindhoven, IEEE, 2015, pp. 1--6.

\bibitem{Dourbois2017PowerTech}
G.~A. Dourbois, P.~N. Biskas, A novel method for the clearing of a day-ahead
  electricity market with mixed pricing rules, in: 2017 IEEE Manchester
  PowerTech, 2017, pp. 1--6.
\newblock \href {http://dx.doi.org/10.1109/PTC.2017.7981100}
  {\path{doi:10.1109/PTC.2017.7981100}}.

\bibitem{iacopoTPRWS2017}
I.~Savelli, A.~Giannitrapani, S.~Paoletti, A.~Vicino, An optimization model for
  the electricity market clearing problem with uniform purchase price and zonal
  selling prices, {IEEE} Trans. Power Syst.\href
  {http://dx.doi.org/10.1109/TPWRS.2017.2751258}
  {\path{doi:10.1109/TPWRS.2017.2751258}}.

\bibitem{sleiszRaisz2017integratedUPP}
{\'A}.~Sleisz, D.~Raisz, Integrated mathematical model for uniform purchase
  prices on multi-zonal power exchanges, Electric Power Systems Research 147
  (2017) 10--21.

\bibitem{UPPOmanual}
{Tabors Caramanis and Associates Inc.},
  \href{www.mercatoelettrico.org/en/MenuBiblioteca/Documenti/20041206UniformPurchase.pdf}{UPPO
  Auction Module User Manual} (2002).
\newline\urlprefix\url{www.mercatoelettrico.org/en/MenuBiblioteca/Documenti/20041206UniformPurchase.pdf}

\bibitem{blanco2014consumer}
R.~Fern{\'a}ndez-Blanco, J.~M. Arroyo, N.~Alguacil, Consumer payment
  minimization under uniform pricing: A mixed-integer linear programming
  approach, Applied Energy 114 (2014) 676--686.

\bibitem{CAISO}
{California Independent System Operator},
  \href{http://www.caiso.com/23cf/23cfe2c91d880.pdf}{{T}echnical {B}ulletin -
  {M}arket optimization details} (2009).
\newline\urlprefix\url{http://www.caiso.com/23cf/23cfe2c91d880.pdf}

\bibitem{Bovo2017}
L.~H. Lam, V.~Ilea, C.~Bovo, Impact of the price coupling of regions project on
  the day-ahead electricity market in italy, in: 2017 IEEE Manchester
  PowerTech, 2017, pp. 1--6.
\newblock \href {http://dx.doi.org/10.1109/PTC.2017.7981215}
  {\path{doi:10.1109/PTC.2017.7981215}}.

\bibitem{schweppe1988spot}
F.~Schweppe, M.~Caraminis, R.~Tabors, R.~Bohn, Spot pricing of electricity,
  Kluwer Academic Publishers, Norwell, MA, 1988.

\bibitem{schweppeCaramanis1982optimal}
M.~C. Caramanis, R.~E. Bohn, F.~C. Schweppe, Optimal spot pricing: Practice and
  theory, {IEEE} Trans. Power App. Syst. PAS-101 (1982) 3234--3245.

\bibitem{GuidelineCACM2015n1222}
{Commission of European Union},
  \href{data.europa.eu/eli/reg/2015/1222/oj}{Commission regulation ({EU})
  2015/1222 of 24 july 2015 establishing a guideline on capacity allocation and
  congestion management} (2015).
\newline\urlprefix\url{data.europa.eu/eli/reg/2015/1222/oj}

\bibitem{bard1998practical}
J.~Bard, Practical bilevel optimization: applications and algorithms, Kluwer
  Academic Press Dordrecht, Netherlands, 1998.

\bibitem{blanco2012unified}
R.~Fern{\'a}ndez-Blanco, J.~M. Arroyo, N.~Alguacil, A unified bilevel
  programming framework for price-based market clearing under marginal pricing,
  {IEEE} Trans. Power Syst. 27~(1) (2012) 517--525.

\bibitem{conejo2012complementarity}
S.~A. Gabriel, A.~J. Conejo, J.~D. Fuller, B.~F. Hobbs, C.~Ruiz,
  Complementarity modeling in energy markets, Vol. 180, Springer Science \&
  Business Media, 2012.

\bibitem{bradley1977applied}
S.~Bradley, A.~Hax, T.~Magnanti, Applied mathematical programming, Addison
  Wesley, 1977.

\bibitem{conejo2010decision}
A.~J. Conejo, M.~Carri{\'o}n, J.~M. Morales, Decision making under uncertainty
  in electricity markets, Vol.~1, Springer, 2010.

\bibitem{blanco2014revenue}
R.~Fern{\'a}ndez-Blanco, J.~M. Arroyo, N.~Alguacil, Revenue-and
  network-constrained market clearing via bilevel programming, in: Proc. of
  Power Systems Computation Conference 2014, IEEE, 2014, pp. 1--7.

\bibitem{blanco2017solution}
R.~Fern{\'a}ndez-Blanco, J.~M. Arroyo, N.~Alguacil, On the solution of
  revenue-and network-constrained day-ahead market clearing under marginal
  pricing {P}art {I}: An exact bilevel programming approach, {IEEE} Trans.
  Power Syst. 32~(1) (2017) 208--219.

\bibitem{binaryGupteConvexHull}
A.~Gupte, S.~Ahmed, M.~S. Cheon, S.~Dey, Solving mixed integer bilinear
  problems using milp formulations, SIAM Journal on Optimization 23~(2) (2013)
  721--744.
\newblock \href {http://dx.doi.org/10.1137/110836183}
  {\path{doi:10.1137/110836183}}.

\bibitem{pereira2005strategic}
M.~V. Pereira, S.~Granville, M.~H. Fampa, R.~Dix, L.~A. Barroso, Strategic
  bidding under uncertainty: a binary expansion approach, {IEEE} Trans. Power
  Syst. 20~(1) (2005) 180--188.

\bibitem{binarizeRoy2007}
J.-S. Roy, “binarize and project” to generate cuts for general
  mixed-integer programs, Algorithmic Operations Research 2~(1).

\bibitem{pyomoBook}
W.~E. Hart, C.~D. Laird, J.-P. Watson, D.~L. Woodruff, G.~A. Hackebeil, B.~L.
  Nicholson, J.~D. Siirola, Pyomo--optimization modeling in python, 2nd
  Edition, Vol.~67, Springer Science \& Business Media, 2017.

\bibitem{cplex2009v12}
{IBM-ILOG}, {CPLEX} {U}ser's {M}anual - {V}ersion 12 {R}elease 7 (2016).

\bibitem{openDAMGitHub}
B.~Corn\'elusse, I.~Savelli, A.~Giannitrapani, S.~Paoletti, A.~Vicino,
  \href{https://github.com/bcornelusse/openDAM}{Open {D}ay-{A}head {M}arket -
  {I}mplementation}, {V}ersion 1.0 (2018).
\newline\urlprefix\url{https://github.com/bcornelusse/openDAM}

\bibitem{openDAMDoc}
B.~Corn\'elusse, I.~Savelli, A.~Giannitrapani, S.~Paoletti, A.~Vicino,
  \href{http://opendam.readthedocs.io}{Open {D}ay-{A}head {M}arket -
  {D}ocumentation} (2018).
\newline\urlprefix\url{http://opendam.readthedocs.io}

\end{thebibliography}

\end{document}